\documentstyle[epsfig]{l-school}
\begin{document}
%
%
\let\ov=\over
\let\lbar=\l
\let\l=\left
\let\r=\right
\def\der#1#2{{\partial{#1}\over\partial{#2}}}
\def\dder#1#2{{\partial^2{#1}\over\partial{#2}^2}}
\def\N{{I\!\!N}}
\def\be{\begin{equation}}
\def\ee{\end{equation}}
%
%
\markboth{S. Bonazzola \& E. Gourgoulhon}{GRAVITATIONAL WAVES FROM NEUTRON
STARS}
%
\setcounter{part}{10}
%
\title{Gravitational waves from neutron stars}
%
\author{S. Bonazzola and E. Gourgoulhon}
\institute{D\'epartement d'Astrophysique Relativiste et de Cosmologie,\\
UPR 176 C.N.R.S.,\\
Observatoire de Paris, \\
F-92195 Meudon Cedex, France}
\maketitle
\section{INTRODUCTION}

A crude estimate of the gravitational luminosity of an object of mass
$M$, mean radius $R$ and internal velocities of order $V$ can be
derived from the quadrupole formula \cite{Blanc96}:
\be     \label{e:quadru,L}
  L \sim {c^5\ov G} \, s^2 \l( {R_{\rm s} \ov R} \r) ^2 
		\l( V \ov c \r) ^6      \ ,
\ee
where $R_{\rm s} := 2 G M / c^2$ is the Schwarzschild radius associated
with the mass $M$ and $s$ is some asymmetry factor: $s=0$ for a
spherically symmetric object and $s\sim 1$ for an object whose shape
is far from that of a sphere. According to formula (\ref{e:quadru,L}),
the astrophysical objects for which 
$s\sim 1$, $R\sim R_{\rm s}$ and $V \sim c$ may radiate a fantastic
power in the form of gravitational waves:
$L\sim {c^5/ G} = 3.6\times 10^{52}$ W, which amounts to
$10^{26}$ times the luminosity of the Sun in the electromagnetic domain!

A neutron star has a radius quite close to its Schwarzschild radius:
$R \sim 1.5 - 3 \, R_{\rm s}$ and its rotation velocity may reach
$V\sim c/2$ at the equator, so that they are a priori valuable
candidates for strong gravitational emission.
The crucial parameter to be investigated is the asymmetry factor $s$.
It is well known that a uniformly
rotating body, perfectly symmetric with respect to its rotation axis
does not emit any gravitational wave ($s=0$).
Thus in order to radiate gravitationally
a neutron star must deviate from axisymmetry. 
P.~Haensel's lecture \cite{Haens96} investigates the deviation from axisymmetry
resulting from irregularities (``mountains'') in the solid crust or from
the neutron star precession. 
In the present lecture, we
investigate two other mechanisms which generate a deviation from
axisymmetry: (i) the spontaneous symmetry breaking resulting from the
development of a triaxial instability in a rapidly rotating neutron
star (\S~\ref{s:symbreak}) and (ii) the distortion induced
by the internal magnetic field of the neutron star (\S~\ref{cw,pulsars}).

\section{SPONTANEOUS SYMMETRY BREAKING} \label{s:symbreak}

A rotating neutron star can spontaneously break
its axial symmetry if the ratio of the rotational kinetic energy
$T$ to the absolute value of the
gravitational potential energy, $|W|$, exceeds some critical value.
This may occur in two different astrophysical situations:
a just born neutron star resulting from a supernova may accrete
matter that has not been ejected by the shock wave, thereby increasing its
kinetic energy; alternatively, an old neutron star in a close binary
system, accreting matter steadily from
its companion, may be spun up until the ratio $T/|W|$ is high enough to
allow the symmetry breaking.

When the critical threshold $T/|W|$ is reached, two kinds of instabilities
may drive the star into the non-axisymmetric state:
\begin{enumerate}
\item the {\em Chandrasekhar-Friedman-Schutz instability}
(hereafter {\em CFS instability})  \cite{Chand70}, \cite{FrieS78},
\cite{Fried78} driven
by the gravitational radiation reaction. 
\item the viscosity driven instability \cite{RobeS63}.
\end{enumerate}
The present lecture puts the accent on the viscosity driven instability.

\subsection{Review of classical results about Maclaurin/Jacobi ellipsoids}
\label{s:Maclaurin}

Let us recall some classical results from the theory
of rotating Newtonian homogeneous bodies. 
It is well known that a self-gravitating incompressible fluid rotating rigidly
at some moderate velocity takes the shape of an axisymmetric ellipsoid: 
the so-called {\em Maclaurin spheroid}. At the critical point 
$T/|W| = 0.1375$ in the 
Maclaurin sequence, two families of triaxial ellipsoids branch
off: the {\em Jacobi ellipsoids} and the {\em Dedekind ellipsoids}.
The former are triaxial ellipsoids rotating rigidly about their
smallest axis in an inertial frame, whereas the latter have a fixed 
triaxial figure in an inertial frame, with some internal fluid circulation
at constant vorticity   
(see ref.~\cite{Chand69} or \cite{Tasso78} for a review of these classical
results).
The Maclaurin spheroids are dynamically unstable for $T/|W|\geq0.2738$.
Thus the Jacobi/Dedekind bifurcation point $T/|W| = 0.1375$ is dynamically 
stable. However, in presence of some dissipative mechanism such as viscosity
or gravitational radiation (CFS instability) that breaks the circulation
or angular momentum conservation, 
the bifurcation point becomes secularly unstable against 
the $l=2, m=2$ ``bar'' mode. Note also that a non-dissipative mechanism such
as the one due to a magnetic field with a component parallel to the rotation 
axis breaks the
circulation conservation \cite{ChrCK96} and may generate
a spontaneous symmetry breaking. 
If one takes into account only the
viscosity, the growth of the bar mode leads to the deformation 
of the Maclaurin spheroid along a sequence of figures close to some Riemann S 
ellipsoids\footnote{The {\em Riemann S} family is formed by 
homogeneous bodies whose 
fluid motion can be decomposed into a rigid rotation about a principal
axis and a uniform circulation whose vorticity is parallel to the rotation
vector. Maclaurin, Jacobi and Dedekind ellipsoids are all special cases
of Riemann S ellipsoids (for more details, cf. Chap.~7 of ref.~\cite{Chand69}
or Sect.~5 of ref.~\cite{LaiRS93}).}   
and whose final state is a Jacobi ellipsoid \cite{PresT73}.
On the opposite, if the gravitational radiation reaction 
is taken into account but not the viscosity, the Maclaurin spheroid evolves
close to another Riemann S sequence towards a Dedekind ellipsoid
\cite{Mille74}.

As we shall see in the next section,
this symmetry breaking  is a particular case of a more general 
phenomenon .

\subsection{Spontaneous breaking of symmetry: a general phenomenon}

\subsubsection{A toy model}

\begin{figure}
\centerline{\epsfig{figure=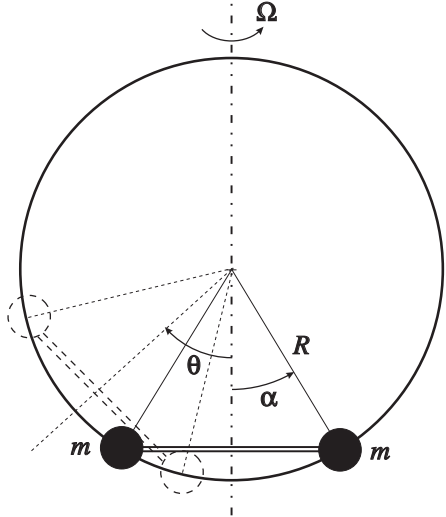,height=7cm}}
\caption[]{\label{f:toy}
Toy model illustrating the phenomenon of 
spontaneous symmetry breaking: the configuration
depicted with dashed lines violates the symmetry with respect to
the rotation axis and
is a stable equilibrium configuration if $\Omega > \Omega_{\rm crit}$.}
\end{figure}

Let us start with a simple example. Consider the toy model of
Fig.~\ref{f:toy} : two heavy points of mass $m$, rigidly connected and
sliding on a circular guide of radius $R$, of negligible mass and
rotating with the angular velocity $\Omega$. 
It is straightforward to write down the expression 
for the potential energy of this system
and to find the equilibrium configuration in a frame
rotating with the guide. With the notations of Fig.~\ref{f:toy},
the potential (potential energy per unit mass) writes:
\be \label{toy1}
	U(\theta) = - (R\Omega)^2 (\sin^2\alpha + \cos2\alpha \, \sin^2\theta )
		  - 2 g R \cos\alpha \cos\theta \ ,     
\ee
$g$ being the gravity acceleration. 
In spite of the invariance of the potential $U$ with respect the
transformation $ \theta \mapsto -\theta$, two equilibium configurations 
exist if the angular velocity is larger than the value
$\Omega_{\rm crit} = [\, g \cos\alpha/(R \cos 2\alpha)\, ]^{1/2} $ :
$\theta=0$ and $\cos\theta= g\cos\alpha/(R \Omega^2 \cos2 \alpha) $.  

This behaviour is typical of the phenomenon of spontaneous symmetry breaking.  
More generally, a dynamical system is said to break
spontaneously its symmetry if there exist solutions
with lower
symmetries than those of the Lagrangian from which the equations of motion
have been derived. 
Other examples of spontaneous symmetry breaking are the phase transitions
(ferromagnetism at the Curie point, melting of solids), the
famous walking stick in Charlie Chapling movies (the stick is axially symmetric
but when the pressure of the hand on its top is larger than
a critical value, the stick bends), the buckling collapse of metallic
structures and so on.

Coming back to the toy model, it is easy to see that, when two equilibrium
configurations exist,  the one is unstable and the other one  is stable. 
For what follows it is more convenient to take
the angular momentum as a free parameter instead of the angular 
velocity.
Let us suppose that some viscous force acts on the two mass points sliding on
the circular guide.  If at time $t=0$, the system is in the equilibrium
configuration $\theta=0 $ and its angular momentum $L$ is larger than
the critical value $ L_{\rm crit}$ corresponding to $\Omega_{\rm crit}$, 
then the system will move to 
the second equilibrium configuration. In this example, we have all the
     main ingredients
     of the spontaneous symmetry breaking mechanism: a parameter describing 
     the equilibrium configuration of the system, a dynamical quantity that is
     not conserved (the kinetic energy) and a total energy that must
     be at its minimum. 
In order to understand the existence of two equilibrium configurations,
it is worth writing the total energy (per unit mass) with respect to the 
inertial frame in terms of the angular momentum $L$:
$E = 1/2 \, L^2 / I + U_{\rm grav}$, 
where $I(\theta)$ is the moment of inertia of the system with respect to the
rotation axis and $U_{\rm grav}$ is the gravitational potential:
$U_{\rm grav} = - 2 g R \cos\alpha\cos\theta$. 
It is clear that, at fixed $L$,  $L^2/I$ decreases when $I$ increases, i.e. when
$\theta$ increases. On the contrary, $U_{\rm grav}$ is an increasing function
of $\theta$. When $L$ is large enough, the first term prevails and $E$ decreases
when $\theta$ increases. 

In order to see the link existing between the symmetry breaking mechanism
of the above toy model and the second order phase transition theory, 
let us briefly recall the Landau theory of these phenomena (see
Chap.~14 of \cite{LandL76}).
In a second order phase transition, a thermodynamical system makes a
continuous transition from a state with a given symmetry to a state with
a lower symmetry. The leading idea in the Landau theory is the expansion
of the thermodynamical potential $\Phi(P,T) $ in a series of a parameter
$\eta $ (the {\em order parameter}) which measures 
how far the thermodynamical system is
from the state of higher symmetry:
\be
	  \Phi(P,T,\eta) = \Phi_0(P,T) + \alpha \eta + A \eta^2
		   +B \eta^3 +C \eta^4 + \cdots
\ee
From invariance considerations, $\alpha$ must be identically equal to zero, 
$B$ must also be equal to zero in order to obtain a stable system. 
On the contrary, $C $ must be positive.
$A$ must be larger than zero in the symmetric phase because the system is 
stable. On the contrary, in the less symmetric phase, values of $\eta$ 
different 
from zero correspond to a stable state. That is possible only if $A < 0 $, 
therefore $A $ must vanish at the critical temperature: $A=a(T-T_c) $. 
In conclusion we can write:
\be \label{Landau}
	    \Phi(P,T,\eta)= \Phi_{0}(P,T) +a(T-T_c)\eta^2 + C\eta^4 \ . 
\ee
The value of $\eta$ in the vicinity of $T_c$ is: 
\be
	       \eta^2= a(T_c-T)/(2C) \ .
\ee
From the expression (\ref{Landau}) for the thermodynamical potential all the
interesting thermodynamical quantities and their jumps at the critical
temperature can be derived.

Coming back to the toy model, consider the total potential
(eq.~(\ref{toy1})) and expand it as function of $\theta $ around the
value $\theta = 0$.  The result is 
\begin{eqnarray}
  U &  = & - (R\Omega\sin\alpha)^2 + R(g\cos\alpha - R\Omega^2 \cos 2\alpha)\  
	\theta^2 \nonumber \\
 & &  + R [ R\Omega^2 \cos 2\alpha - (g\cos\alpha)/4 ] \  \theta^4/3 
	+ O(\theta^6) \ . \label{toy2}
\end{eqnarray}
It is easy to see the analogy between expressions
(\ref{Landau}) and (\ref{toy2}). In fact $\theta$ is the order
parameter of the toy model: when $\theta=0 $
the system is axisymmetric and it breaks this symmetry for $\theta >0 $.
The potential $U $ is the analogue of the thermodynamical
potential $\Phi$.  No extra terms with odd power of $\theta $
appear, there is a critical value
of $\Omega $ for which the quadratic terms vanish and, last but
    not least, the quartic term is positive in the vicinity of 
$\Omega_{\rm crit}$. It is straightforward to
    define a ``temperature'' of the system. Let us put $T=1/(R\Omega)^2$;
then $T_c=\cos2\alpha \, / (g R \cos\alpha)$, 
$a= 2 (g R \cos\alpha)^3 / \cos^2 2\alpha$ and
$C = (g R \cos\alpha)^2 / \cos 2\alpha$. 
The reader can push the analogy further by defining the
equivalent ``thermodynamical quantities'' and compute their
discontinuities at the transition point. 

An equivalent treatment (but slightly more complicated) can be employed 
by replacing $ \Omega $ by the angular momentum $L$. In this case 
$\alpha$ must be different from zero, in order to have a finite angular
momentum when $ \theta=0 $.

\subsubsection{The case of rotating stars}

The mechanism of axial symmetry breaking for a rotating star
is very similar to the mechanism discribed in the toy model.
When the angular momentum $L$ of the star is larger than a critical value 
$ L_c $, the symmetry of the star changes in order to increase its moment of
inertia $I$ and consequently to decrease its rotational kinetic energy for
a given angular momentum $L$. 

In what follows, it is more convenient to parametrize the rotation of
the star by the ratio $R$ of the rotational kinetic energy to
the potential energy: $ R = T / |W| $ instead of the angular momentum $L$.
In fact, the critical value of $ R_c $ at which the symmetry
is broken depends only on the equation of state (hereafter EOS) 
of the fluid which
constitutes the star. Moreover, in the case of a polytropic EOS, 
($ P = \alpha n^\gamma $, 
$n$ and $P$ being the density and the pressure respectively) $ R_c $ depends
on $ \gamma $ only. As recalled in \S~\ref{s:Maclaurin}, 
for an incompressible star  $ R_c = 0.1375 $ (secular instability).
 
The symmetry breaking process is, of course, more complicated than the
one described for the toy model. The number of degrees of freedom
of the star is infinite instead of being just one for the toy model, 
consequently the analogy cannot be pushed further.

In order to understand what follows, the following theorem is needed:
\begin{quote}
{\bf Theorem:} Among all the possible angular velocity distributions 
in a steady state configuration, rigid rotation 
($ \Omega = {\rm const.}$) minimizes the energy of the star (see the problem at
the end of this lecture). 
\end{quote}
We may now understand what happens to an axisymmetric
rigidly rotating star when the angular velocity is slightly larger than the
critical one. Consider for the moment the case without dissipative
mechanisms. 
Since at constant angular momentum --- larger than the critical one ---
the configuration which minimizes the total energy of the {\em rigidly} rotating 
star is a triaxial one, the star tends to break the axial symmetry and starts 
to deform to evolve to the new equilibrium configuration. 
In absence of viscosity or any other dissipative mechanisms,
the angular momentum of each shell of the star is conserved and the angular
velocity is no more uniform. It turns out that this new configuration, 
slightly non-axisymmetric has a total energy larger than the old one because
the rotation is no more rigid (cf. the above theorem). Therefore
the star does not evolve in this way and stays in the axisymmetric 
configuration.

We may say that the conservation of the angular momentum of each shell
acts as a potential barrier that inhibits the transition to the non-axisymmetric 
equilibrium configuration.
The situation is completely different when viscosity is present. 
When the star starts to deform, the viscosity spreads out the angular 
momentum, so that the rotation becomes rigid and the star can go to the 
non-axisymmetric equilibrium configuration. It is obvious that the time 
taken to make the transition depends on the time required to rigidify 
the motion: the smaller the viscosity,
the longer the time to make the transition. For this reason
this instability is called {\em secular}.
When the angular momentum is larger than a second critical value 
$L_{\rm c,dyn}$ the potential barrier due to the non-rigid rotation can be 
overcome 
and the transition toward the new equilibrium configuration is possible.
The corresponding critical value of $R$ is $ R_{\rm c,dyn} = 0.2738 $ for an
incompressible fluid. 
This instability is called {\em dynamical} for it occurs on a dynamical 
timescale. 

The instability described above is at the origin of the transition of the 
star from a Maclaurin spheroid to a triaxial Jacobi ellipsoid. 
Since viscosity-driven evolution preserves the angular momentum but not 
the kinetic energy, the final equilibrium configuration has the same mass, 
the same angular momentum and rigid rotation. 
If the total mechanical energy is conserved but the conservation of the 
angular momentum is violated, then the Maclaurin steady state configuration
makes a transition toward the triaxial Dedekind ellipsoid, which is not 
rigidly rotating. 

Actually, there exists many other bifurcation points, all of them
being parame\-trized by a given value of the $ T/W $ ratio 
(pear configuration found by Poincar\'e, fission bifurcation, etc...)


From what has been said, the analogy between the mechanism of spontaneous 
symmetry breaking of the toy model and that of the rotating star should 
have clearly appeared.
Consequently, the transition between the Maclaurin and the Jacobi ellipsoids
can be studied by using the theory of second order phase transition, as we
have done for the toy model. Historically, 
Bertin and Radicati \cite{BertR76} were the first 
to recognize in 1976 that the bifurcations in the steady state configurations
of rotating homogeneous stars are genuine second order phase transitions.
More recently, Christodoulou, Kazanas, Shlosman and Tohline \cite{ChKST95a}
have systematically used this analogy to study the different bifurcations of 
rotating homogeneous stars and to investigate the possibility of a 
transition from one steady state configuration to another one. 
Not only have they found new results
but they have also given more insight to the understanding of this mechanism.
We highly recommend their articles 
\cite{ChKST95a}, \cite{ChKST95b}, \cite{ChKST95c}, \cite{ChKST95d}
to the interested reader.

\subsection{Previous results for compressible Newtonian stars}

The results on Maclaurin spheroids recalled above have been extended
to compressible fluids, modeled by a polytropic EOS,  
by a number of authors. First of all Jeans \cite{Jeans19}, \cite{Jeans28}
has shown that a bifurcation point
towards triaxial configurations can exist only if the adiabatic index $\gamma$
is larger than $\gamma_{\rm crit}\simeq2.2$. The interpretation is that the EOS
must be stiff enough for the bifurcation point to occur at an angular velocity
lower than the maximum angular velocity $\Omega_{\rm K}$
for which a stationary solution exists. $\Omega_{\rm K}$ is reached when
the centrifugal force exactly balances the gravitational force at the
equator of the star, and for this reason is called the {\em Keplerian velocity};
if the star were forced to rotate at $\Omega > \Omega_{\rm K}$, it would
lose some matter from the equator. By numerical calculations, 
James \cite{James64} has refined Jeans' result to
\be  \label{e:gam,crit}
	\gamma_{\rm crit} = 2.238 \qquad \mbox{(James 1964)}.
\ee

As concerns the secular instability in the compressible case, Ipser \&
Managan \cite{IpseM85} have shown that the $m=2$ Jacobi-like
bifurcation point has the same location along uniformly rotating
sequences as the $m=2$ Dedekind-like point, as in the incompressible
case.  For $\gamma<\gamma_{\rm crit}$,
the CFS instability still exists for modes $l=m\geq3$ \cite{Manag85},
\cite{ImaFD85} but not the viscous instability: if the viscosity is
important, its effect is always stabilizing by acting against the CFS
instability.  Lindblom \cite{Lindb95} has shown that for a star of
$1.5\, M_\odot$ constructed by means of a $\gamma=2$ polytropic EOS, the
CFS instability is suppressed at temperatures $T<5\times 10^6$ K by
the shear viscosity and at $T>10^{10}$ K by the bulk viscosity.  These
results have been confirmed by Yoshida \& Eriguchi \cite{YoshE95} (see
also Sect.~2 of ref.~\cite{FrieI92} and references therein).  For
completeness, let us note that if the neutron star interior is superfluid,
the CFS instability is suppressed by the ``mutual friction'' of its
components \cite{LindM95}.

Regarding the gravitational wave signal from rotating neutron stars that
undergo the above triaxial instabilities, Ipser \& Managan
\cite{IpseM84} have
examined the case of polytropic stars with $\gamma=2.66$ and $\gamma=3$
which have bifurcated along a triaxial Jacobi-like sequence, under the effect
of the viscosity driven instability. Wagoner \cite{Wagon84}
has computed the
gravitational signal from an accreting neutron star --- modeled by
nearly spherical homogeneous objects --- for the five lowest modes of
the CFS instability. Recently, Lai \& Shapiro \cite{LaiSh95}
have determined the
gravitational wave form from newborn neutron stars --- modeled as
self-similar ellipsoids --- undergoing the bar mode $l=m=2$ of the CFS
instability. 

\subsection{Generation of gravitational waves}

It is worth discussing 
the role played by the two instability mechanisms, CFS and viscosity-driven,
in the emission of gravitational waves by a rotating neutron star. 
Let us consider a just born neutron star (from the core collapse of 
a massive star) with a kinetic energy $ T $ larger than the critical
one. For zero viscosity, the star breaks its symmetry via the CFS mechanism.
Its fate is a triaxial Dedekind-like ellipsoid. Because the principal axes
of this ellipsoid are fixed in an inertial frame the quadrupole moment 
is constant and the gravitational radiation stops.
On the contrary if the viscosity is larger than a 
critical value \cite{LindD77}, the CFS instability is inhibited and 
the star evolves toward a Jacobi configuration; 
it loses energy and angular momentum via gravitational
radiation until its kinetic energy achieves the critical value 
$ T_c $. It turns out that this second mechanism is more efficient for 
gravitational wave generation
than the first one. In the intermediate case, the 
gravitational radiation light curve depends on the ratio between the 
rising time of the two instabilities. The real life is more complicated: 
in fact the EOS is far from being a
polytropic. Consequently, the conditions for a bar instability
depend on the EOS {\em and} on the mass of the star. Moreover, neutron stars
are strongly relativistic objects and post-Newtonian or even 
post-post-Newtonian approximations are not sufficient
to describe their steady states of rotation.  
A fortiori, the results obtained in the Newtonian theory can hardly be used to 
predict the gravitational radiation from these objects.

In what follows, we shall try to answer to the following questions:
\begin{enumerate}
\item Among all the EOS proposed for nuclear matter, are there some 
   stiff enough to allow the bar instability ?
\item What is the influence of the general relativistic effects on the bar
   instability ?
\item In the case of a positive answer to the first question, and under the 
   hypothesis that the general relativistic effects do not kill the bar
   instability, what are the masses of the stars for which the bar instability
   can develop~?
\item Are these masses compatible with the masses of neutron stars
 already observed ?
\end{enumerate}

\subsection{Finding the equilibrium configurations of a rotating star in 
the Newtonian regime} \label{s:equil,rot,Newt}

In order to show how the Newtonian results can be extended to 
general relativity, we need to explain how the equilibrium configurations of
rotating barotropic stars are computed. Therefore, this section may
appear quite technical. The reader not interested in mathematical
and technical problems, can skip this section and go directly
to section \ref{resu,poly}. 
He should only remember that the exact solution for the motion
of a triaxial rotating star in the framework of general relativity 
does not exist yet,
the results presented in the next sections being obtained
under adequate approximations. 

Consider a steady state star rotating with constant angular velocity 
 $\Omega $ about the $ z $ axis. In a frame rotating with
the same angular velocity and about the same rotation axis, the star will 
appear static, therefore
the problem of finding a steady state configuration becomes a problem
of finding a static configuration. Let $ n $, $P(n) $ be the
density and the pressure of the fluid respectively. The barotropic 
EOS $P=P(n)$ is supposed to be known. The equilibrium equations reads
\begin{eqnarray} 
& &1/n\ \partial P(n)/\partial \rho +\partial U/\partial\rho
 - \Omega^2 \rho = 0 \label{equilibre1} \\
& & 1/n\  \partial P(n)/\partial z + \partial U/\partial z = 0 \ , 
					\label{equilibre2}
\end{eqnarray}
where $ U(x,y,z) $ is the self-gravitational potential, and $ \rho^2=
x^2+y^2 $.
By introducing the specific enthalpy of the fluid, 
\be \label{eos}
 H(n)=  \int {1\over n} {dP\over dn} dn 
\ee
the system of differential equations (\ref{equilibre1})-(\ref{equilibre2})
can be integrated immediately to yield
\be \label{int-premiere}
	  H + U - {1\ov 2} \Omega^2 \rho^2 = {\rm const.}
\ee
Hence once $U$ is known and $ H $ is fixed at the 
centre of
the star, $ H $ and therefore $ n $ can be easily computed at all the points
of the star. The surface of the star is defined by $ H =0 $.

The potential $ U $ must be computed by solving in a consistent way the 
2-D Poisson  equation
\be \label{Poisson}
 \Delta U = 4 \pi G \, n  
\ee
where $G$ is the gravitational constant.  
The integration of the system of equations (\ref{Poisson}), 
(\ref{int-premiere}) and (\ref{eos}) is performed by 
relaxation: take a trial distribution of matter $ n $, compute the
potential $U$ by solving the Poisson equation, compute the corresponding
$H $ by means of Eq.~(\ref{int-premiere}), inverse the relationship 
$H(n) $ to obtain
the density $ n $ and solve the Poisson equation to find
a new potential $ U $ until convergence is achieved. About 50
iterations are necessary to obtain numerical solutions that differ less
than $ 10^{-12} $ between two different iterations. 
In a such a way  a 2-D equilibrium configuration is obtained. 

Once the equilibrium configuration is found, its stability can be studied.
For this purpose we proceed in the following way: 
instead of stopping the relaxation
at the $ J^{\rm th}$ iteration say, the procedure is continued
altering slightly the gravitational potential at the $ (J+1)^{\rm th} $ 
iteration: some perturbation 
$ \epsilon\,  r^2 P_2(\cos \theta) \cos(2\phi)  $ is added to 
the axisymmetric potential,  where
$ r^2 = \rho^2 + z^2 $, $ P_2$ is the Legendre polynomial of degree 2 and 
 $ \epsilon $
a numerical constant ($ \epsilon=10^{-6} $). At the iteration $ J+2 $ the 
perturbation is
switched off and the relaxation continues. If the equilibrium configuration
is stable, the perturbed solution relaxes to the axisymmetric solution;
if not, it relaxes to the triaxial equilibrium configuration. In such a way,
it is possible to find the critical value of $ \gamma$, $T/|W|$,  etc...

\subsection{Extension to general relativity}

The analysis of the triaxial instability of rotating stars has been
recently extended to general relativity \cite{BonFG96}
in the case of {\em rigid} rotation. This latter assumption, crucial
for finding a first integral of motion for non-axisymmetric relativistic
configurations
by following the procedure of Carter \cite{Carte79}, corresponds
physically to the viscosity driven instability, but not to the CFS
instability. 
The translation of the technique presented in \S~\ref{s:equil,rot,Newt}
from the Newtonian framework to the
general relativistic framework is not completely straightforward. 
Before considering the relativistic analysis let us consider again 
the Newtonian description of triaxial configurations. 

\subsubsection{Analysis of Newtonian triaxial configurations} 
\label{s:Newt,3D}
Let us first come back to the problem of finding axisymmetric 
stationary solutions of rigidly rotating stars in Newtonian theory. 
In this theory two approaches are 
possible: The first one consists in finding an  
{\em equilibrium} configuration in a {\em non-Galilean} rotating frame
(cf. \S~\ref{s:equil,rot,Newt}). 
In this frame the velocity vanishes and the term $ \Omega^2 \rho $ in 
Eq.~(\ref{equilibre1}) is the apparent acceleration  (centrifugal
acceleration). 

In the second approach (Galilean approach), the chosen reference frame is
Galilean, the velocity of
the fluid $v_{(\phi)} = \Omega \rho $ is not zero, and the term $ \Omega \rho^2$
is due to the term $v^k \nabla_k v^i $ of the Euler equation 
(inertial acceleration). In both approaches the problem consists in finding
a steady-state solution. Note that the mass conservation equation
\be \label{contin} 
\partial n/\partial t + \nabla_i(nv^i) = 0 
\ee
is identically satisfied and that the equation (\ref{Poisson})
for the gravitational potential $ U $ is the same one in the two approaches.

The situation is different when looking for triaxial solutions.
In fact for a Jacobi-like configuration, a steady state (equilibrium) 
solution exists only in the rotating frame.
The questions that naturally arise are the following ones:
is it possible to find an equivalent ``equilibrium'' configuration in
a Galilean frame ? How to solve the Euler equation ?
In order to answer these questions consider the unchanged Poisson
equation (\ref{Poisson}) and the Euler and mass conservation equations 
($ x^1=\rho,x^2=z, x^3 = \phi $)
\begin{eqnarray}  
& &\partial v_i/\partial t + v^k \nabla_k v_i + 
\partial H(\rho,z,\phi,t)/\partial x^i+
\partial U(\rho,z,\phi,t)/\partial x^i=0 \\
& &\partial n(\rho,z,\phi,t)/\partial t + 
v^k\partial n(\rho,z,\phi,t)/\partial x^k +n \nabla_k v^k =0
\label{e:Mass,conserv}
\end{eqnarray}
altogether with the Poisson equation (\ref{Poisson}) for the potential 
which is unchanged.
Let us look for a solution corresponding to a triaxial star 
undergoing rigid rotation about the $z$ axis:
$ v_{(\rho)} =0,\ v_{(z)} =0,\ v_{(\phi)} = \Omega \rho $. 
The equations for the velocity read:
\begin{eqnarray}
& & - \Omega^2 \rho+
\partial H(\rho,z,\phi,t)/\partial \rho +
\partial U(\rho,z,\phi,t)/\partial \rho =0  \label{e:Euler,newt,1} \\
& &
\partial H(\rho,z,\phi,t)/\partial z +
\partial U(\rho,z,\phi,t)/\partial z =0 \\
& &
\partial H(\rho,z,\phi,t)/\partial \phi +
\partial U(\rho,z,\phi,t)/\partial \phi=0 \ , \label{e:Euler,newt,3}
\end{eqnarray}
whereas Eq.~(\ref{e:Mass,conserv}) reduces to
\be \label{conti2}
\partial n(\rho,z,\phi,t)/\partial t +
\Omega \, \partial n(\rho,z,\phi,t)/\partial \phi =0
\ee
It is easy to see that Eqs.~(\ref{e:Euler,newt,1})-(\ref{e:Euler,newt,3}) 
are satisfied if
$H(\rho,z,\phi,t) + U(\rho,z,\phi,t) - 1/2 \, \Omega^2 \rho^2 = C(t) $,
where $ C(t) $ is an arbitrary function of $ t $.
This relation is a first integral of motion, 
very similar to relation (\ref{int-premiere}).

The continuity equation (\ref{conti2}) can be satisfied if we 
take $C(t) = {\rm const}$
and consider $ n$, $H$, and $U $ as functions of the
new variables $(\rho,z,\psi)$ where 
\be \label{e:def,psi}
  \psi := \phi - \Omega t \ .
\ee 
Eq.~(\ref{conti2}) then becomes $-\Omega \, \partial n/\partial \psi + 
\Omega \, \partial n/\partial \psi = 0$, i.e. is
identically satisfied.

From a geometrical point of view, we may say that the 
Newtonian spacetime of a rotating triaxial star 
has a one-parameter symmetry group, whose
trajectories (orbits) are defined by $\psi = {\rm const}$. 
The spacetime vectors that generate this symmetry group are called 
{\em Killing vectors}. 
In the frame associated with a Killing vector, 
the star appears in a steady state. 
The use of the notion of symmetry group or Killing vector fields
allows the
absolute definition (independent of any reference frame) of a steady
state configuration. This is the definition that is going to be used in
general relativity. 

\subsubsection{Rigid motion in general relativity} \label{s:rigid,GR}

Before the symmetry breaking, the spacetime generated by the rotating 
star can be considered as {\em stationary} and {\em axisymmetric}, which
means that there exist two Killing vector fields, $k^\alpha$ and $m^\alpha$,
such that $k^\alpha$ is timelike (at least far from the star) and $m^\alpha$
is spacelike and its orbits are closed curves. Moreover, in the case
of rigid rotation, the spacetime is {\em circular}, which means that 
the 2-spaces orthogonal to both $k^\alpha$ and $m^\alpha$ are integrable in
global 2-surfaces \cite{Carte73}. This latter property considerably
simplifies the study of rotating stars because some global coordinates
$(t,r,\theta,\phi)$ may be chosen so that the metric tensor components
exhibit only one non-vanishing off-diagonal term ($g_{t\phi}$).
$t$ and $\phi$ are coordinates associated with respectively the
Killing vectors $k^\alpha$ and $m^\alpha$: $k^\alpha = \partial/\partial t$
and
$m^\alpha = \partial/\partial \phi$. The remaining coordinates $(r,\theta)$
span the 2-surfaces orthogonal to both $k^\alpha$ and $m^\alpha$.
In all numerical work on rotating stars to date (see e.g.
ref.~\cite{BoGSM93} for a review) {\em isotropic coordinates} are chosen,
for which the
two-dimensional line element differs from the flat space one by a
conformal factor $A^2$. In these coordinates, the components of the metric
tensor are given by
\be \label{e:g:axisym}
g_{\alpha\beta}\, dx^\alpha dx^\beta = - N^2 \, dt^2 +
   B^2 r^2\sin^2\theta (d\phi - N^\phi\, dt)^2
   + A^2 \l[ dr^2 + r^2 \, d\theta^2 \r]\ ,
\ee
where the four functions $N$, $N^\phi$, $A$ and $B$ depend on the
coordinates $(r,\theta)$ only, the coordinates $(t,\phi)$ being
associated with the Killing vector fields.  

When the axisymmetry of the star is broken, the stationarity of spacetime
is also broken. As discussed above, in the Newtonian theory, there is no
inertial (Galilean) frame in which a rotating triaxial object appears 
stationary,
i.e. does not depend upon the time. It can be stationary only in a 
corotating frame, which is not inertial, so that the stationarity is
broken in this sense. 
In the general relativistic
case, where the notion of a global inertial frame is in general meaningless,
a rotating triaxial system is not stationary for it radiates away
gravitational waves. Even if a corotating frame could be defined,
the body could not be in a steady state in this frame, because it 
loses energy and angular momentum via gravitational radiation. 

However, at the very point of the symmetry breaking, no gravitational
wave has been emitted yet. As an approximation we neglect any 
subsequent gravitational radiation. Then, for a
rigid rotation, there exists one Killing vector field $l^\alpha$, which is
proportional to the fluid 4-velocity $u^\alpha$, hence
\be \label{e:u=lamb*l}
	u^\alpha = \lambda \, l^\alpha                                  \ ,
\ee
where $\lambda$ is a strictly positive scalar function.
Equation (\ref{e:u=lamb*l}) is the definition of a {\em rigid motion}
according to Carter \cite{Carte79} \footnote{More generally the 
relativistic concept of {\em rigidity} is defined by the
requirement that the {\em expansion tensor} 
[cf. Eq.~(\ref{e:expansion,tens}) below]
associated with the vector field $u^\alpha$ should vanish; it is easy
to show that this requirement is fulfilled if $u^\alpha$ takes the form
(\ref{e:u=lamb*l}).}:
 (i) there exists a Killing vector field;
(ii) the fluid 4-velocity is parallel to this Killing vector.
In the axisymmetric and stationary case, the rigid motion corresponds
to the constant angular velocity $\Omega := u^\phi / u^t$, the
Killing vector entering equation (\ref{e:u=lamb*l}) being then
\be  \label{e:l(k,m)}
	l^\alpha = k^\alpha + \Omega \, m^\alpha \ ,
\ee
where $k^\alpha$ and $m^\alpha$ are the two Killing vectors defined above.
The constancy of $\Omega$ ensures that $l^\alpha$ is a Killing vector too.
The proportionality constant $\lambda$ of equation (\ref{e:u=lamb*l}) is
nothing else than the component $u^t$ of the 4-velocity $u^\alpha$,
where $t$ is the coordinate associated with the Killing vector $k^\alpha$.
Note that the Killing vector $l^\alpha$ is generally timelike close to the
star and spacelike far from it (beyond the ``light-cylinder'').

In the non-axisymmetric case, $k^\alpha$ and $m^\alpha$ can no longer be
defined as Killing vectors. We make instead the assumption that there exist
(i) a vector field $k^\alpha$ which is timelike at least far from the star,
(ii) a vector field $m^\alpha$, which commutes with $k^\alpha$, is spacelike
everywhere and whose field lines are closed curves, 
(iii) a constant $\Omega$ such that the vector $l^\alpha$ defined by
the combination (\ref{e:l(k,m)}) is a Killing vector and
(iv) the fluid 4-velocity $u^\alpha$ is parallel to $l^\alpha$.
These hypotheses
are the geometric translation of the approximation of rigid rotation and
negligible gravitational radiation. The commutativity of $k^\alpha$ and
$m^\alpha$ ensures that a coordinate system $(t,r,\theta,\phi)$
can be found such that $k^\alpha = \partial/\partial t$ and
$m^\alpha = \partial/\partial \phi$.

\subsection{First integral of fluid motion in general relativity}

The purpose of this section is to show that in the framework
of general relativity,
there still exists a first integral for the fluid motion, this result
being exact in the axisymmetric and stationary case, and being valid within
the approximation of the existence of the Killing vector $l^\alpha$ in the
3-D case. 

As detailed in ref.~\cite{BonFG96}, the first integral can be found
in an elegant geometrical way, using the {\em canonical form} of the
equation of fluid dynamics introduced by Lichnerowicz \cite{Lichn67}
and Carter \cite{Carte79} and involving Cartan's calculus on differential
forms. However, in the present lecture, we follow a different approach, 
based on the formulation of relativistic
hydrodynamics in an accelerated frame (Appendix~\ref{a:hydro}). In doing so,
a parallel can be developed
with the two points of view of Newtonian physics recalled in
\S~\ref{s:Newt,3D} (that of a rotating observer and an inertial 
(Galilean) observer).

The rotating observer is the observer ${\cal O}_{\rm F}$ comoving with the 
fluid, whose 4-velocity is
$u^\alpha$. The relativistic generalization of the ``fixed'' inertial
observer is chosen to be the Eulerian observer ${\cal O}_{\rm E}$
of the 3+1 formalism \cite{AbraY96}, whose 4-velocity is denoted by
$n^\alpha$. 

\subsubsection{The point of view of the comoving observer}

Let us consider the Euler equation (\ref{e:Euler,DV}) in the frame of
the fluid observer ${\cal O}_{\rm F}$. Using notations of
Appendix~\ref{a:hydro},
one has in this case, $v^\alpha = u^\alpha$, $V^\alpha = 0$ and $E=e$, so
that Eq.~(\ref{e:Euler,DV}) reduces to 
\be  \label{e:Euler,comov}
  {1\ov e+p} \overline\nabla_\alpha p +
	 a_\alpha = 0   \ ,
\ee
with $a_\alpha = u^\mu \nabla_\mu u_\alpha$. Let us recall that 
$\overline\nabla_\alpha$ denotes the covariant derivative within the 
3-dimensional plane orthogonal to ${\cal O}_{\rm F}$ wordlines. 
 Using Eq.~(\ref{e:u=lamb*l})
and the Killing identity  $\nabla_{(\alpha} l_{\beta)} = 0$,
we get $a_\alpha = - \overline\nabla_\alpha \ln\lambda$. 
Eq.~(\ref{e:Euler,comov}) then becomes
\be
   {1\ov e+p} \overline\nabla_\alpha p 
      - \overline\nabla_\alpha \ln\lambda = 0 \ ,
\ee
from which a first integral of motion is immediately obtained:
\be \label{e:int,prem,OF}
    H - \ln\lambda = {\rm const.}
\ee
where $H$ is defined by
\be \label{e:H,def,rel}
   H := \int  {dp \ov e+p} = \ln \l( {e+p\ov m_{\rm B} \, n} \r) \ ,
\ee
$m_{\rm B}$ being the mean baryon mass. The second equality in
Eq.~(\ref{e:H,def,rel}) is a consequence of the First Law of Thermodynamics
at zero temperature. This definition of $H$ constitutes a relativistic
generalization of the specific enthalpy introduced by Eq.~(\ref{eos}).

\subsubsection{The point of view of the Eulerian observer}

Let us now consider the Eulerian observer ${\cal O}_{\rm E}$.
His 4-velocity is
\be
    n^\alpha = {1\ov N} \l( k^\alpha + N^\alpha \r)  \ ,
\ee
where $N$ is the so-called {\em lapse function} and
$N^\alpha$ the {\em shift vector} \cite{AbraY96}. In the present case,
$N^\alpha = N^\phi \, m^\alpha$. Using
notations of Appendix~\ref{a:hydro}, one has $v^\alpha = n^\alpha$,
$\omega_{\alpha\beta}=0$ (the world lines of ${\cal O}_{\rm E}$ are
orthogonal to the hypersurfaces $t={\rm const}$)\footnote{the fact that
the rotation 2-form of the Eulerian observer vanishes, explains why
${\cal O}_{\rm E}$ is sometimes called the {\em locally non-rotating observer}
\cite{Barde70}.}, $\theta_{\alpha\beta} = - K_{\alpha\beta}$,
where $K_{\alpha\beta}$ is the extrinsic curvature tensor of the
hypersurfaces $t={\rm const.}$ \cite{AbraY96},
$V^\alpha = N^{-1} \, (\Omega - N^\phi)\, m^\alpha = N^{-1}
 \, l^\alpha - n^\alpha$,
$\Gamma = N u^t = N \lambda$.
The Euler equation (\ref{e:Euler,vdV}) then becomes
\begin{eqnarray}
  & & n^\mu \nabla_\mu V^\alpha - a_\mu V^\mu n^\alpha +
  V^\mu \overline\nabla_\mu V^\alpha   -
	K^\alpha_{\ \, \mu}  V^\mu
  + (K_{\mu\nu} V^\mu V^\nu - a_\mu V^\mu) V^\alpha \nonumber \\
 & & \qquad  + {1\ov E+p} \l( \overline\nabla^\alpha p +
	V^\alpha n^\mu \nabla_\mu p \r) + a^\alpha = 0 \label{e:Euler,OE} \ .
\end{eqnarray}
After straightforward calculations, making use of the fact that $l^\alpha$
is a Killing vector, the various terms that appear in
this equation can be expressed as
\begin{eqnarray}
  & & n^\mu \nabla_\mu V^\alpha = -{1\ov N} \l( n^\mu \nabla_\mu \ln N \r)
      l^\alpha - {1\ov N} K^\alpha_{\ \, \mu} l^\mu \\
  & & V^\mu \overline\nabla_\mu V^\alpha = - \Gamma^{-3} 
	\overline\nabla^\alpha \Gamma
  + (\Gamma^{-2} - 1) \overline\nabla^\alpha \ln N 
	+ n^\mu \nabla_\mu \ln N \, V^\alpha
	\nonumber  \\
  & & \qquad + 2 N^{-1} \, K^\alpha_{\ \, \mu} l^\mu 
	+ N^{-2} \, K_{\mu\nu} l^\mu l^\nu
	\, n^\alpha \\
  & & a^\alpha = \overline\nabla^\alpha \ln N \\
  & & K_{\mu\nu}\!V^\mu V^\nu = N^{-2}\!K_{\mu\nu} l^\mu l^\nu = 
	- \Gamma^{-3} n^\mu \nabla_\mu \Gamma + (\Gamma^{-2} - 1) 
	n^\mu \nabla_\mu \!\ln N \ 
\end{eqnarray}
Accordingly, Eq.~(\ref{e:Euler,OE}) becomes
\be \label{e:Euler,OE,1}
    \nabla_\alpha ( H + \ln N - \ln \Gamma ) + N^{-1} 
	n^\mu \nabla_\mu ( H + \ln N - \ln \Gamma ) \, l_\alpha = 0 \ .
\ee
Taking the scalar product with $l^\alpha$ and using the fact that
$l^\alpha$ is a symmetry generator, leads to 
$n^\mu \nabla_\mu ( H + \ln N - \ln \Gamma ) = 0$. Introducing this latter
relation in (\ref{e:Euler,OE,1}), we get 
\be \label{e:int,prem,OE}
	H + \ln N - \ln \Gamma = {\rm const.} 
\ee
This first integral of motion is exactly the same as that obtained in the
comoving frame [Eq.~(\ref{e:int,prem,OF})] (remember that 
$\lambda = \Gamma / N$). In the Newtonian limit,
$\ln N \rightarrow U$ and $\ln \Gamma \rightarrow \rho^2 \Omega^2 /2$,
so that Eq.~(\ref{e:int,prem,OE}) reduces to Eq.~(\ref{int-premiere}), 
as expected. 

\subsection{Gravitational field equations}

Let us now say a few words about the equations for the gravitational field
in the general relativistic case. For axisymmetric rotating stars, these
equations are well known (see e.g. ref.~\cite{BoGSM93}). 
As concerns the 3-D case and 
within the approximation made above (existence of the Killing vector 
$l^\alpha$), 
a natural approach would be to use the Geroch formalism \cite{Geroc71} which
reduces the 4-dimensional Einstein equations to 3-dimensional equations
by forming the quotient of spacetime by the trajectories of the Killing 
vector field. 
However in the present case, the Killing vector $l^\alpha$ is timelike
inside the light cylinder and spacelike outside it. Consequently the
second-order operators that appear in Geroch's formalism change from  
elliptic to hyperbolic type across the light cylinder. From the numerical point
of view, we would rather have operators of a constant kind. For this reason,
we do not consider Geroch formalism but instead the classical 3+1 formalism
\cite{AbraY96},
i.e. the foliation of spacetime by spacelike hypersurfaces $\Sigma_t$. 
Within $\Sigma_t$, we choose coordinates $(r,\theta,\phi)$ according to 
the prescription of \S~\ref{s:rigid,GR}. Next we introduce the coordinate
$\psi$ as in Eq.~(\ref{e:def,psi}) : $\psi := \phi -\Omega t$.
All the metric coefficients are then functions of the three variables
$(r,\theta,\psi)$. In the equations of the 3+1 formalism (cf. \S~3.2 of
\cite{AbraY96}) appear partial
derivatives with respect to $t$, $r$, $\theta$ and $\phi$. We replace
them by partial derivatives with respect to $r$, $\theta$ and $\psi$ 
according to the rule
\be
    \der{u}{t} \longrightarrow - \Omega \der{u}{\psi}
	\qquad 
    \der{u}{\phi} \longrightarrow \der{u}{\psi} \ .
\ee
We then obtain only 3-dimensional equations. Another simplification
arises from considering only the dominant terms in the non-axisymmetric
part of the equations. We report to ref.~\cite{BonFG96} for further details. 
Let us simply mention that within our approximation, the triaxial metric
tensor writes
\begin{eqnarray} \label{e:g:3D}
  g_{\alpha\beta}\, dx^\alpha\, dx^\beta & = &
   - N(r,\theta,\psi)^2\, dt^2  
   + B(r,\theta,\psi)^2  \, 
     r^2\sin^2\theta \l[ d\phi - N^\phi(r,\theta) \, dt \r] ^2 
						\nonumber \\ 
   & & + A(r,\theta,\psi)^2 
	 \l[ dr^2 + r^2 \, d\theta^2 \r]\ .     
\end{eqnarray}
Note that the metric coefficients given by Eq.~(\ref{e:g:3D})
are the components of the metric tensor with respect to the coordinates
$(t,r,\theta,\varphi)$ and expressed as functions of the coordinates
$(r,\theta,\psi)$.

Within this approximation, the Einstein equations reduce to four elliptic
equations, for the functions $N$, $N^\phi$, $AN$ and $BN$. In the 
Newtonian limit, the equation for $N$ reduces to the Poisson equation
(\ref{Poisson}).

\subsection{Numerical results}

The non-linear 3-D elliptic equations resulting from the formalism 
developed above are solved iteratively
by means of a spectral method \cite{BoGSM93}, \cite{Gourg96}.
The initial conditions are 
stationary axisymmetric configurations, constructed by a 2-D general
relativistic numerical code \cite{BoGSM93}, \cite{SaBGH94}. 
The initial lapse function $N$ is perturbed by a non-axisymmetric
$l=2, m = \pm 2$ ``bar'' term, in the same way as the Newtonian gravitational
potential in \S~\ref{s:equil,rot,Newt}. The stability is determined by 
examining the subsequent evolution.

\begin{figure}
\centerline{
\epsfig{figure=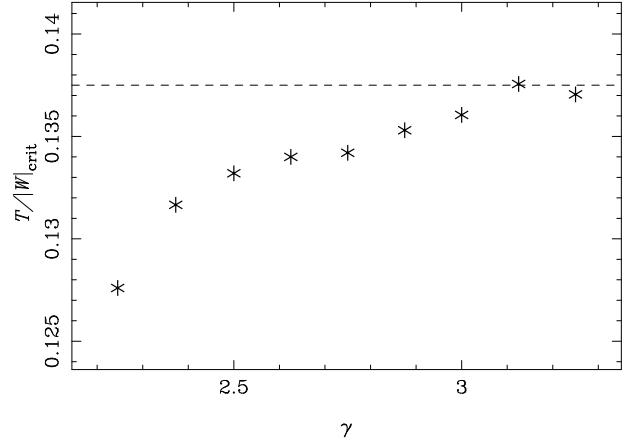,height=7cm}
}
\caption[]{\label{f:T/W-gam}
Ratio of the kinetic energy $T$ to the gravitational potential energy $W$
at the triaxial Jacobi-like bifurcation point along a sequence of
rotating Newtonian polytropes, as a function of the adiabatic index $\gamma$.
The dashed horizontal line corresponds to the theoretical value of $T/|W|$ for
incompressible Maclaurin spheroids.}
\end{figure}

\subsubsection{Tests of the numerical code}

As a test of the relativistic code,
the same value of the
critical adiabatic index as that found by James \cite{James64}
(Eq.~\ref{e:gam,crit}) has been obtained at the Newtonian limit.
Also in the Newtonian regime, it has been verified that when the
polytropic index $\gamma$ increases, the ratio $T/|W|$ at which the
triaxial instability occurs, tends to the classical value for
incompressible fluids: $T/|W|_{\rm crit}(\gamma=\infty)=0.1375$
(cf. \S~\ref{s:Maclaurin}), as shown in Fig.~\ref{f:T/W-gam}.
Another test of the code consists in comparing results in the compressible
polytropic case with previous numerical calculations, in the non-relativistic
regime.
Ipser \& Managan \cite{IpseM81} and Hachisu \& Eriguchi \cite{HachE82}
have obtained numerical
models of Newtonian triaxial rotating polytropes, analogous to the Jacobi
ellipsoids. They did not determine $\gamma_{\rm crit}$\footnote{However,
Ipser \& Managan state that their results indicate that the
critical adiabatic index lies somewhere in the range $2.22\leq \gamma_{\rm crit}
\leq2.28$} but performed calculations with fixed $\gamma \geq 2.66$.
The location of the triaxial bifurcation
point along a sequence of $\gamma=3$ polytropes resulting from the
relativistic code \cite{BonFG96} used at the Newtonian limit has been
compared with that obtained
by the above authors. The agreement is better than 0.5\% with the 
critical angular velocity of Hachisu \& Eriguchi \cite{HachE82}
and of the order
of 2\% with their critical value of $T/|W|$; with Ipser \& Managan
\cite{IpseM81}, the agreement is better than 0.5\% on both quantities.

\begin{figure}
\centerline{
\epsfig{figure=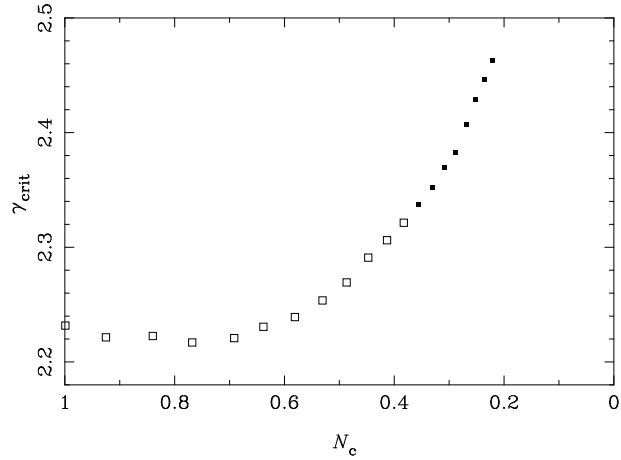,height=7cm}
}
\caption[]{\label{f:gcrit-Nc}
Critical polytropic index $\gamma_{\rm crit}$ as function of the lapse
function
$N_{\rm c}$ measured at the centre of the star. Black boxes indicate
configurations
unstable with respect to radial oscillations.}
\end{figure}

\subsubsection{Results for polytropes} \label{resu,poly}

The investigation of relativistic polytropic stars is interesting in several
respects. First, it represents a natural extension of former classical
works restricted to the Newtonian case. Second, a polytropic EOS
does not suffer from the thermodynamical
 inconsistency relative to
tabulated EOS (cf. Sect.~4.2 of ref.~\cite{SaBGH94}).
It provides therefore an approximate but consistent model for real 
stars which allows a first investigation of relativistic effects.
Since Newtonian polytropes obey a scaling law, $\gamma_{\rm crit}$,
the critical index for which the inset of the secular triaxial instability
coincides with the maximum rotating case, is a global constant.
In the relativistic case however, relativistic effects are
supposed to influence the symmetry breaking and the critical index will depend
on an appropriate parameter measuring the relativistic character of the object.
A well suited quantity is the value $N_{\rm c}$ of the {\em lapse
function} \cite{AbraY96} at the centre of the star.
We thus obtain a 2--dimensional parameter
space for maximum rotating stars which is intersected
by a curve representing the metastable configurations and thus separating
the regions of stable and unstable stars respectively.
Figure~\ref{f:gcrit-Nc} shows the dependence of
$\gamma_{\rm crit}$ on $N_{\rm c}$ ranging from the Newtonian to the 
extreme relativistic regime.
In the moderately relativistic domain there appears a very slight decrease of
 $\gamma_{\rm crit}$. The negative
slope reveals that the inset of relativistic effects tends to destabilize
the star (within the limitations imposed by the approximate character of the
theoretical approach \cite{BonFG96}),
the maximum decrease of $\gamma_{\rm crit}$ being about $0.6\,\%$.
In the strong field region we observe a smooth
growth of $\gamma_{\rm crit}$ beyond the maximum mass configurations
due to the now persistently increasing stabilizing relativistic effects.

\begin{table}
\centerline{
\begin{tabular}{lllllll}
\hline\noalign{\smallskip}
  $\displaystyle{{\rm EOS}\atop \ }$    &  
  $\displaystyle{{M_{\rm max}^{\rm stat}\atop [M_\odot]}}$      &
  $\displaystyle{{M_{\rm max}^{\rm rot}\atop [M_\odot]}}$       &
  $\displaystyle{{P_{\rm K}\atop [{\rm ms}]}}$  &
  $\displaystyle{{P_{\rm break}\atop [{\rm ms}]}}$      &
  $\displaystyle{{H_{\rm c,break}\atop \ }}$    &
  $\displaystyle{{M_{\rm break}\atop [M_\odot]}}$       \\
 \noalign{\smallskip}
\hline\noalign{\smallskip}
  HKP     & 2.827 & 3.432 & 0.737 & 1.215 & 0.161 & 1.80 \\
  WFF2    & 2.187 & 2.586 & 0.505 & 0.755 & 0.30 & 1.951 \\
  WFF1    & 2.123 & 2.528 & 0.476 & 0.728 & 0.27 & 1.736 \\
  WGW     & 1.967 & 2.358 & 0.676 & 
		\multicolumn{3}{c}{marginally stable} \\
  Glend3  & 1.964 & 2.308 & 0.710 & \multicolumn{3}{c}{stable} \\
  FP      & 1.960 & 2.314 & 0.508 & 0.604 & 0.465 & 1.736 \\
  DiazII  & 1.928 & 2.256 & 0.673 & \multicolumn{3}{c}{stable} \\
  BJI     & 1.850 & 2.146 & 0.589 & \multicolumn{3}{c}{stable} \\
  WFF3    & 1.836 & 2.172 & 0.550 & 0.714 & 0.325 & 1.909 \\
  Glend1  & 1.803 & 2.125 & 0.726 & \multicolumn{3}{c}{stable} \\
  Glend2  & 1.777 & 2.087 & 0.758 & \multicolumn{3}{c}{marginally stable} \\
  PandN   & 1.657 & 1.928 & 0.489 & \multicolumn{3}{c}{stable} \\
\noalign{\smallskip}
\hline\noalign{\smallskip}
\end{tabular}
}
\caption[]{\label{t:resu,EOS}
Neutron star properties according to various EOS: $M_{\rm max}^{\rm stat}$
is the maximum mass for static  configurations, 
$M_{\rm max}^{\rm rot}$ is the maximum mass for rotating stationary 
configurations, $P_{\rm K}$ is the corresponding Keplerian period,
$P_{\rm break}$ is the rotation period below which the symmetry breaking occurs, 
$H_{\rm c, break}$ is the central log-enthalpy at the bifurcation point 
and $M_{\rm break}$ is the corresponding gravitational mass.  
The EOS are ordered by decreasing values of $M_{\rm max}^{\rm stat}$.
}
\end{table}

\subsubsection{Results for realistic equations of state}

We have determined in what condition
the symmetry breaking may occur for rapidly rotating neutron stars
built upon twelve EOS resulting from nuclear physics calculations. These
``realistic'' EOS are the same as those used in ref.~\cite{SaBGH94}
and we refer to this paper for a description of each EOS. 
The EOS are labeled by the following abbreviations: PandN refers to the
pure neutron EOS of Pandharipande \cite{Pandh71}, 
BJI to model IH of Bethe \& Johnson \cite{BethJ74}, 
FP to the EOS of Friedman \& Pandharipande \cite{FrieP81}, 
HKP to the  $n_0 = 0.17\ {\rm fm}^{-3}$ model of Haensel et al. \cite{HaeKP81},
DiazII to model II of Diaz Alonso \cite{Diaz85}, 
Glend1, Glend2 and Glend3 to respectively the case 1, 2, and 3 of 
Glendenning EOS \cite{Glend85},
WFF1, WFF2 and WFF3 to respectively the 
${\rm AV}_{14}+{\rm UVII}$, ${\rm UV}_{14}+{\rm UVII}$ and 
${\rm UV}_{14}+{\rm TNI}$ models of Wiringa et al. \cite{WirFF88}, 
and WGW to the $\Lambda_{\rm Bonn}^{00}+{\rm HV}$ model of Weber et al.
\cite{WebGW91}.   

Our results are shown in table~\ref{t:resu,EOS}. For a given EOS, the
axisymmetric rotating models form a two parameter family; each model can be
labeled by its central energy density $e_{\rm c}$ and its (constant)
angular velocity $\Omega$. For a given value of $e_{\rm c}$, $\Omega$
varies from zero to the Keplerian velocity $\Omega_{\rm K}$. 
Following the method described above, we have
searched for a symmetry breaking of configurations rotating at the
Keplerian velocity. 
For five EOS (Glend1, Glend3, DiazII, BJI and PandN), no symmetry 
breaking was found, whatever the value of $\Omega_{\rm K}$. These EOS are
listed as ``stable'' in table~\ref{t:resu,EOS}. For two
EOS (WGW and Glend2) the evolution 
was not conclusive. These EOS are listed as ``marginally stable'' in
table~\ref{t:resu,EOS}. A better numerical precision could lead to a
definitive conclusion. For five EOS (HKP, FP, WFF1, WWF2 and WFF3), 
the bar mode reveals to be instable for some Keplerian velocities. 
Table~\ref{t:resu,EOS} gives the period $P_{\rm break}$, 
gravitational mass $M_{\rm break}$ and central log-enthalpy 
$H_{\rm c,break}$ (cf. Eq.~(\ref{e:H,def,rel})) of the configuration having
the lowest angular velocity and for which the symmetry breaking occurs.

\section{CW EMISSION FROM PULSARS} \label{cw,pulsars}

As stated in the introduction, 
rapidly rotating neutron stars (pulsars) might be an important source of
continuous gravitational waves. Morever, the expected gravitational frequency 
is related to the rotation frequency and lies in the frequency bandwidth of
the forthcoming LIGO and VIRGO interferometric detectors. 
While \S~\ref{s:symbreak} considers strong asymmetries resulting from
an instability, we here investigate permanent slight asymmetries and the
resulting continuous wave (CW) emission. 

Various kinds of pulsar asymmetries have been suggested in the literature:
first the crust of a neutron star is solid, so that its shape may not 
necessarily be axisymmetric under the effect of rotation, as it would
be for a fluid: deviations from axisymmetry are supported by anisotropic
stresses in the solid. The shape of the crust does not depend only on the 
early history of the neutron star, especially on the phase of 
crystalization of the crust\cite{Haens96}, but also on star quakes.
Due to its violent formation (supernova) or due to its environment
(accretion disk), the
rotation axis may not coincide with a principal axis of the neutron
star moment of inertia and the star may precess. 
Even if it keeps a perfectly axisymmetric shape, a freely precessing body
radiates gravitational waves. This effect is discussed in P.Haensel's lecture
\cite{Haens96}. We consider another source of asymmetry here, linked to the
magnetic field. Indeed, 
neutron stars are known to have important magnetic fields and the
magnetic pressure (Lorentz forces exerted on the conducting
matter) can distort the star 
if the magnetic axis is not aligned with the rotation axis, which is 
widely supposed to occur in order to explain the pulsar phenomenon.

\begin{figure}
\centerline{
\epsfig{figure=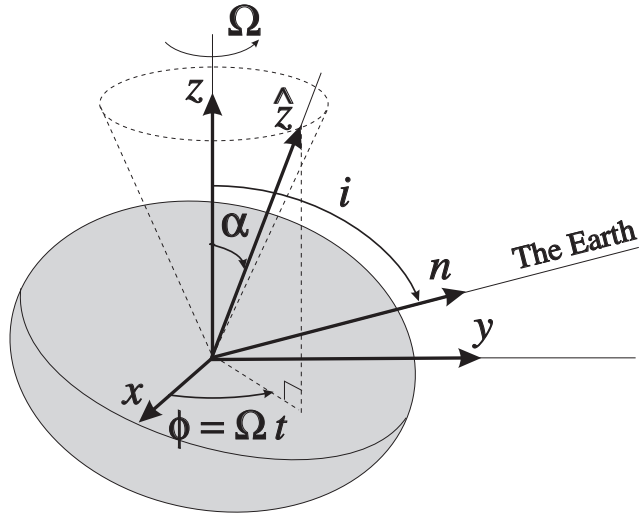,height=7cm}
}
\caption[]{\label{f:etoile}
Geometry of the distorted neutron star.}    
\end{figure}

\subsection{Generation of gravitational waves by a rotating star}
\label{s:generation} 

Let us consider a neutron star rotating at the angular velocity
$\Omega$ about some axis (cf. Fig.~\ref{f:etoile}). In the framework of general 
relativity, the mass quadrupole moment relevant for gravitational
radiation is {Thorne's quadrupole moment} ${\cal I}_{ij}$ \cite{Thorn80}. 
Ipser \cite{Ipser71} has shown that 
the leading term in the gravitational radiation field $h_{ij}$ is  
given by a formula which is structurally identical to the 
quadrupole formula for Newtonian sources (cf. \S~2.2 of L. Blanchet's
lecture \cite{Blanc96}), the Newtonian quadrupole being simply replaced
by Thorne's quadrupole. The non-axisymmetric deformation of neutron stars
being very tiny, the total Thorne's quadrupole can be linearly decomposed
into the sum of two pieces:
$ {\cal I}_{ij} = {\cal I}_{ij}^{\rm rot} + {\cal I}_{ij}^{\rm dist} $,
where ${\cal I}_{ij}^{\rm rot}$ is the quadrupole moment due to rotation
(${\cal I}_{ij}^{\rm rot} = 0$ if the configuration is static) and
${\cal I}_{ij}^{\rm dist}$ is the quadrupole moment due to the
process that distorts the star, for example an internal magnetic field
or anisotropic stresses from the nuclear interactions.
Let us make the assumption that the distorting process has a
privileged direction, i.e. that two of the three 
eigenvalues of ${\cal I}_{ij}^{\rm dist}$ are equal. Let then $\alpha$ be the
angle between the rotation axis and the principal axis of 
${\cal I}_{ij}^{\rm dist}$ which corresponds to the eigenvalue which is not
degenerate (cf. Fig.~\ref{f:etoile}). The two modes $h_+$ and $h_\times$
of the gravitational radiation field in a transverse traceless gauge
are then given by \cite{BonaG96}
\begin{eqnarray}
   h_+ & = & h_0 \sin\alpha \Big[
	{1\ov 2} \cos\alpha\sin i\cos i \cos\Omega(t-t_0) \nonumber \\
	& & \qquad \qquad
 - \sin\alpha {1+\cos^2 i\ov 2} \cos2\Omega (t-t_0) \Big] \label{e:h+,gen} \\
   h_\times & = & h_0 \sin\alpha \Big[
	{1\ov 2} \cos\alpha\sin i\sin\Omega(t-t_0) \nonumber \\
	& & \qquad \qquad
	- \sin\alpha \cos i \sin2\Omega (t-t_0) \Big] \ , \label{e:hx,gen}
\end{eqnarray}
where $i$ is the inclination angle of the ``line of sight'' with respect
to the rotation axis (see Fig.~\ref{f:etoile}) and
\be \label{e:h0,eps}
    h_0 = {16\pi^2 G\ov c^4} {I\, \epsilon\ov P^2\, r} \ , 
\ee
where $r$ is the distance of the star,
$P=2\pi/\Omega$ is the rotation period of the star, 
$I$ its moment of inertia with respect of the rotation axis and
$ \epsilon := - {3/2} \, {{\cal I}_{\hat z\hat z}^{\rm dist} /  I} $    
the {\em ellipticity} resulting from the distortion process. 

From the formul\ae\ (\ref{e:h+,gen})-(\ref{e:hx,gen}), it is clear that
there is no gravitational emission if the distortion axis is aligned
with the rotation axis ($\alpha=0\ \mbox{or}\ \pi$). If both axes are
perpendicular ($\alpha=\pi/2$), the gravitational emission is monochromatic
at twice the rotation frequency. In the general case ($0<|\alpha|<\pi/2$),
it contains two frequencies: $\Omega$ and $2\Omega$. For small values
of $\alpha$ the emission at $\Omega$ is dominant.
Replacing the physical
constants by their numerical values results in the following expression 
for the gravitational wave amplitude [Eq.~(\ref{e:h0,eps})]
\be \label{e:h0,num}
    h_0 = 4.21\times 10^{-24} \ \Big[ {{\rm ms}\ov P} \Big] ^2
	\Big[ {{\rm kpc}\ov r} \Big] 
	\Big[ {I\ov 10^{38} {\ \rm kg\, m}^2} \Big]
	\Big[ {\epsilon \ov 10^{-6} } \Big] .
\ee
Note that $I=10^{38} {\ \rm kg\, m}^2$ is a representative value for
the moment of inertia of a $1.4 \, M_\odot$ neutron star (see Fig.~12
of ref.~\cite{ArneB77}).
	
For the Crab pulsar, $P=33 {\ \rm ms}$ and $r=2{\ \rm kpc}$, so that
Eq.~(\ref{e:h0,num}) becomes
\be
   h_0^{\rm Crab} = 1.89\times 10^{-27}  
	\Big[ {I\ov 10^{38} {\ \rm kg\, m}^2} \Big]
	\Big[ {\epsilon \ov 10^{-6} } \Big] .
\ee
For the Vela pulsar, $P=89 {\ \rm ms}$ and $r=0.5{\ \rm kpc}$, hence
\be
   h_0^{\rm Vela} = 1.06\times 10^{-27}  
	\Big[ {I\ov 10^{38} {\ \rm kg\, m}^2} \Big]
	\Big[ {\epsilon \ov 10^{-6} } \Big] .
\ee
For the millisecond pulsar\footnote{We do not consider the
``historical'' millisecond pulsar PSR 1937+21 for it is more than twice
farther away.} PSR 1957+20, $P=1.61 {\ \rm ms}$ and $r=1.5{\ \rm kpc}$, hence
\be \label{e:h0,1957+20}
   h_0^{\mbox{\tiny 1957+20}} = 1.08\times 10^{-24}  
	\Big[ {I\ov 10^{38} {\ \rm kg\, m}^2} \Big]
	\Big[ {\epsilon \ov 10^{-6} } \Big] .
\ee
At first glance, PSR 1957+20 seems to be a much more favorable candidate than
the Crab or Vela. However, in the above formula, $\epsilon$ is in units of 
$10^{-6}$ and the very low value of the period derivative $\dot P$ of
PSR 1957+20 implies that its $\epsilon$ is at most $2\times 10^{-9}$
\cite{Haens96}, \cite{NeCJT95}. 
Hence the maximum amplitude one can expect for this
pulsar is $h_0^{\mbox{\tiny 1957+20}} \sim 1.7 \times 10^{-27}$ and
not $1.08\times 10^{-24}$ as Eq.~(\ref{e:h0,1957+20}) might suggest. 

\subsection{The specific case of magnetic field induced deformation} 
\label{s:magnetic}

Let us assume that the distortion of the star is due to its own magnetic field. 
It is then expected that the ellipticity $\epsilon$ is proportional to 
the square of the {\em magnetic dipole moment} $\cal M$ of the 
neutron star \cite{BonaG96}:
\be \label{e:eps(M)}
    \epsilon = \beta \, {\mu_0 \ov 4\pi} { R^2\ov G \, I^2} \, {\cal M}^2\ .
\ee
In this formula, $R$ is the circumferential equatorial radius of the star
and $\beta$ is a dimensionless coefficient which 
 measures the efficiency of this
magnetic structure in distorting the star. In the following, we shall call 
$\beta$ the {\em magnetic distortion factor}. 
Provided the magnetic field amplitude does not take (unrealistic) huge values
($> 10^{14}$ T), the formula (\ref{e:eps(M)}) is certainly true, even if 
the magnetic field
structure is quite complicated, depending on the assumed
electromagnetic properties of the fluid: normal conductor, superconductor,
ferromagnetic... 

Now the observed spin down of radio pulsars is very certainly due
to the low frequency magnetic dipole radiation. $\cal M$ is then linked to the
observed pulsar period $P$ and period derivative $\dot P$ by
(cf e.g. Eq.~(6.10.26) of ref.~\cite{Strau84})
\be \label{e:M(Ppoint)}
    {\cal M}^2 = {4\pi \ov \mu_0} {3 c^3\ov 8 \pi^2} 
		{I P \dot P\ov \sin^2\alpha} \ ,
\ee
where $\alpha$ is the angle between the magnetic dipole moment 
{\boldmath $\cal M$} and the
rotation axis. For highly relativistic configurations, the vector 
{\boldmath $\cal M$} is defined in the weak-field near zone (cf. Sect.~2.5
of ref.~\cite{BoBGN95}), so is $\alpha$.
Inserting Eqs.~(\ref{e:eps(M)}) and (\ref{e:M(Ppoint)}) 
into  Eq.~(\ref{e:h0,eps}) leads to the gravitational wave amplitude
\be \label{e:th0,def}
  h_0 = 6\beta \, {R^2 \dot P\ov c r P \sin^2\alpha} \ ,
\ee
which can be cast in a numerically convenient form:
\be \label{e:th0,num}
    h_0 = 6.48\times 10^{-30} \, {\beta\ov \sin^2\alpha} \, 
	\Big[ {R\ov 10 {\, \rm km}} \Big] ^2
	\Big[ {{\rm kpc}\ov r} \Big] 
	\Big[ {{\rm ms}\ov P} \Big]
	\Big[ {\dot P\ov 10^{-13}} \Big] \ .
\ee

Among the 706 pulsars of the catalog by Taylor et al. \cite{TayML93},
\cite{TaMLC95}, the
highest value of $h_0$ at fixed $\alpha$, $\beta$ and $R$, as given by 
Eq.~(\ref{e:th0,num}), is achieved by the Crab pulsar
($P=33 {\ \rm ms}$, $\dot P = 4.21\times 10^{-13}$, 
$r=2{\ \rm kpc}$), followed by Vela ($P=89 {\ \rm ms}$,
$\dot P = 1.25\times 10^{-13}$, $r=0.5{\ \rm kpc}$)
and PSR~1509-58 ($P=151 {\ \rm ms}$,
$\dot P = 1.54\times 10^{-12}$, $r=4.4{\ \rm kpc}$):
\begin{eqnarray} 
   h_0^{\rm Crab} & = & 4.08\times 10^{-31}   
	\Big[ {R\ov 10 {\, \rm km}} \Big] ^2 \, {\beta\ov \sin^2\alpha}
					 \label{e:th0,Crab}  \\
   h_0^{\rm Vela} & = & 1.81\times 10^{-31}  
	\Big[ {R\ov 10 {\, \rm km}} \Big] ^2 \, {\beta\ov \sin^2\alpha}
					 \label{e:th0,Vela}\\
   h_0^{\mbox{\tiny 1509-58}} & = & 1.50\times 10^{-31}  
	\Big[ {R\ov 10 {\, \rm km}} \Big] ^2 \, {\beta\ov \sin^2\alpha} \\
   h_0^{\mbox{\tiny 1957+20}} & = & 4.51\times 10^{-37}  
	\Big[ {R\ov 10 {\, \rm km}} \Big] ^2 \, {\beta\ov \sin^2\alpha} \ .
\end{eqnarray}
We have added the millisecond pulsar
PSR 1957+20 ($P=1.61 {\ \rm ms}$,
$\dot P = 1.68\times 10^{-20}$, $r=1.5{\ \rm kpc}$) considered in 
\S~\ref{s:generation} to the list. 
From the above values, it appears that 
PSR 1957+20 is not a good candidate. This is not
suprising since it has a small magnetic field (yielding a low $\dot P$). 
Even for the Crab and Vela pulsars, which have a large $\dot P$, 
the $h_0$ values as given by Eqs.~(\ref{e:th0,Crab}), 
(\ref{e:th0,Vela}) are, at first glance, not very encouraging. Let us
recall that with the $10^{-22} \ {\rm Hz}^{-1/2}$ expected sensitivity of 
the VIRGO experiment at a frequency of 30 Hz \cite{BrilG96}, \cite{Bondu96}, 
the minimal amplitude detectable within three years of integration is 
\be \label{e:hmin}
	h_{\rm min} \sim 10^{-26} \ . 
\ee
Comparing this number with Eqs.~(\ref{e:th0,Crab})-(\ref{e:th0,Vela}), 
one realizes that in order to 
lead to a detectable signal, the angle $\alpha$ must be small and/or
the distortion factor $\beta$ must be large. In the former case, the emission 
occurs mainly at the frequency $\Omega$. 
From Eq.~(\ref{e:th0,def})
the gravitational wave amplitude can even be arbitrary large if 
$\alpha\rightarrow 0$. However, if $\alpha$ is too small, let say 
$\alpha < 10^{-2}$, the simple magnetic braking formula (\ref{e:M(Ppoint)})
certainly breaks down. So one cannot rely on a tiny $\alpha$ to yield a 
detectable amplitude. The alternative solution is to have a large $\beta$. 
For a Newtonian incompressible fluid with a uniform magnetic field, 
$\beta=1/5$ \cite{BonaG96}. In the following section, we give the 
$\beta$ coefficients computed for
more realistic models (compressible fluid, realistic equation
of state, general relativity taken into account) with various 
magnetic field distributions.

\subsection{Numerical results} \label{s:mag,num,res}

We have developed a numerical code to compute the deformation of 
magnetized neutron stars within general relativity \cite{BonaG96}. 
This code is an extension of that presented in ref.~\cite{BoBGN95}.   
We report to this latter reference for details about the 
relativistic formulation of Maxwell
equations and the technique to solve them. Let us simply recall here
that the solutions obtained are fully relativistic and 
self-consistent, all the effects of the
electromagnetic field on the star's equilibrium (Lorentz force, spacetime
curvature generated by the electromagnetic stress-energy) being taken into
account. The magnetic field is axisymmetric and poloidal.  
The numerical technique is based on a spectral method \cite{Gourg96}, 
\cite{BoGSM93}. 

\begin{figure}
\centerline{
\epsfig{figure=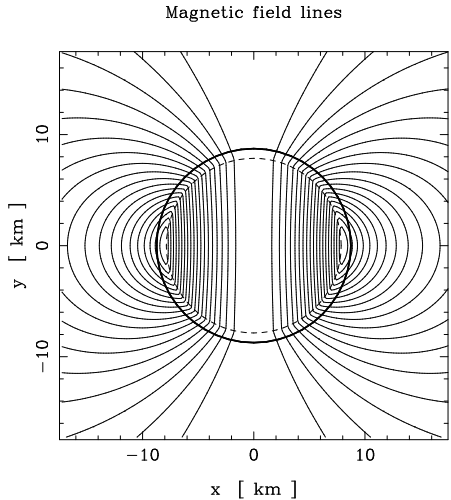,height=7cm}
}
\caption[]{\label{f:mag:j:croute}
Magnetic field lines generated by a current distribution localized in 
the crust of the star. The thick line denotes the star's surface
and the dashed line the internal limit of the electric current distribution.  
The distortion factor corresponding to this situation is $\beta=8.84$.}
\end{figure}

The reference (non-magnetized) configuration is taken to be a $1.4\, M_\odot$
static neutron star built with
the equation of state ${\rm UV}_{14}+{\rm TNI}$ 
of Wiringa, Fiks \& Fabrocini \cite{WirFF88}. 
This latter is a modern, medium stiff equation of state. 
The circumferential radius is
$R=10.92\ {\rm  km}$, the baryon mass $1.56 \, M_\odot$,
the moment of inertia $I=1.23\times 10^{38} {\ \rm kg\, m}^2$
and the central value 
of $g_{00}$ is 0.36, which shows that such an object is highly relativistic. 
Various magnetic field configurations have been considered; the most 
representative of them are presented hereafter.

Let us first consider the case of a perfectly conducting interior (normal
matter, non-superconducting).  
The simplest magnetic configuration compatible with the MHD equilibrium of the star 
(cf. ref~\cite{BoBGN95})
results in electric currents in the
whole star with a maximum value at half the stellar radius in the equatorial plane. 
The computed
distortion factor is $\beta=1.01$, which is above the $1/5$ value of the 
uniform magnetic field/incompressible fluid Newtonian model \cite{BonaG96} 
but still very low. 

Another situation corresponds to electric currents localized in the neutron
star crust only. Figure~\ref{f:mag:j:croute} presents one such configuration: 
the electric current is limited to the zone $r>r_*= 0.9 \, r_{\rm eq}$. 
The resulting distortion factor is $\beta=8.84$. 

\begin{figure}
\centerline{
\epsfig{figure=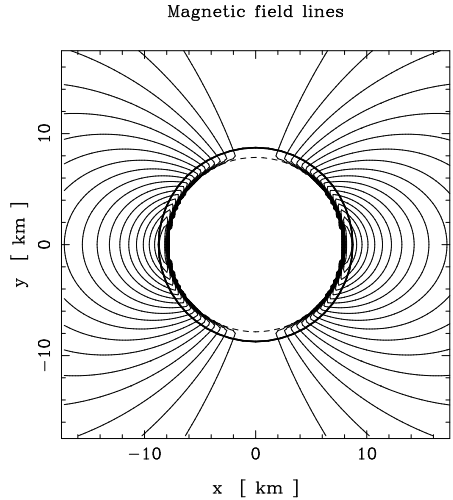,height=7cm}
}
\caption[]{\label{f:mag:supra}
Magnetic field lines generated by a current distribution exterior to a 
type I superconducting core. The thick line denotes the star's surface
and the dashed line the external limit of the superconducting region. 
The distortion factor corresponding to this situation is $\beta=157$.}
\end{figure}

In the case of a superconducting interior, of type I, which 
means that all magnetic field has been expulsed from the superconducting 
region, the distortion factor somewhat increases.
In the configuration depicted in Fig.~\ref{f:mag:supra}, the
neutron star interior is superconducting up to $r_*=0.9\, r_{\rm eq}$. 
For $r>r_*$, the matter is assumed to be a perfect conductor carrying some
electric current. The resulting
distortion factor is $\beta=157$. 
For $r_*=0.95\, r_{\rm eq}$, $\beta$ is even higher: $\beta = 517$. 

The above values of $\beta$, of the order $10^2 - 10^3$, though much higher than
in the simple normal case, are still too low to 
lead to an amplitude detectable by the first generation of 
interferometric detectors in the case of the Crab or Vela pulsar 
[cf. Eqs.~(\ref{e:th0,Crab}), (\ref{e:th0,Vela}) and (\ref{e:hmin})]. 
It is clear that the more disordered the magnetic field the higher $\beta$,
the extreme situation being reached by a stochastic magnetic
field: the total magnetic dipole moment $\cal M$ almost vanishes, in 
agreement with the observed small value of $\dot P$, whereas the mean value 
of $B^2$ throughout the star is huge. 
Note that, according to Thompson \& Duncan \cite{ThomD93}, turbulent dynamo
amplification driven by convection in the
newly-born neutron star may generate small scale magnetic fields as strong as
$3\times 10^{11}{\ \rm T}$ with low values of $B_{\rm dipole}$ outside the
star and hence a large $\beta$. 

In order to mimic such a stochastic magnetic field, we
have considered the case of counter-rotating electric currents.
The resulting distortion factor can be as high as 
$\beta=5.7\times 10^3$. 
 
If the neutron star interior forms a type II superconductor, 
the magnetic field inside the star
is organized in an array of quantized magnetic flux tubes, each tube containing
a magnetic field $B_{\rm c} \sim 10^{11} \ {\rm T}$ \cite{Ruder91}.
As discussed by Ruderman \cite{Ruder91}, the crustal stresses induced by the pinning of 
the magnetic flux tubes is of the order $B_{\rm c} B / 2\mu_0$, 
where $B$ is the mean value of the magnetic
field in the crust ($B\sim 10^8 \ {\rm T}$ for typical pulsars). This
means that the crust is submitted to stresses $\sim 10^3$ higher than in
the uniformly distributed magnetic field 
(compare $B_{\rm c} B/2\mu_0$ with $B^2/2\mu_0$).
The magnetic distortion factor $\beta$ should increase in the same
proportion. We have not done any numerical computation to confirm this
but plan to study type II superconducting interiors in a future work.

\section{DISCUSSION AND CONCLUSION}

\subsection{Spontaneous symmetry breaking}

From the results presented in Table~\ref{t:resu,EOS}, 
it appears that only neutron stars
whose mass is larger than
$1.74\, M_\odot $ meet the conditions of spontaneous symmetry breaking via
the viscosity-driven instability. The above minimum mass is much lower 
than the maximum mass
of a fast rotating neutron star for a stiff EOS ($3.2\ M_\odot $ \cite{SaBGH94}). 
Note that the critical period at which
the instability happens ($P=1.2$ ms) is not far from the lowest observed one
($1.56$ ms).  The question that naturally arises
is : do these heavy neutron stars exist in nature ? 
Only observations can give the
answer; in fact, the numerical modelling of a supernova
core and its collapse \cite{Mulle96} cannot yet provide us with a reliable
answer.
The masses of 17 neutron stars (all in binary systems) are known \cite{ThAKT93}. 
Among them, four masses (all in binary radio pulsars) are known 
with a precision better than $10\% $  and they turn out to be around
$1.4\ M_\odot $ (see Table III in A.~Wolszczan's lecture \cite{Wolsz96}). 
Among the X-ray binary neutron stars, two of them seem to have a higher mass:
4U 1700-37 and Vela X-1 
($1.8\pm 0.5\ M_\odot $ and $1.8\pm 0.3\ M_\odot $ respectively). 
These objects show that neutron stars in binary systems may
have a mass larger than  $1.7\ M_\odot  $.

A natural question that may arise is: why do X-ray binary neutron stars,
which are believed to be the progenitors
of binary radio pulsars, would have a mass larger than the latter ones ?
We have not yet any reliable answer to this question. 
A first (pessimistic) answer is
that the measurements of X-ray neutron star masses are not as reliable 
(compare the error bars
of the masses of the binary radio pulsars with the ones of the X-ray binaries
in Fig.~3 of ref.~\cite{ThAKT93}). Actually it should be noticed that 
the error bars of the X-ray pulsars do not have the
same statistical meaning as the error bars of the binary radio pulsars
\cite{Lindb96}:
they give only the extremum limits of neutron star masses
in the X-ray binary. Consequently $1.4\ M_\odot $ is not incompatible with
these masses. If this is really the case forget all we have said about
the above instability mechanisms: only the CFS mechanism for $m > 2 $ can 
work, provided that the viscosity is low enough, which does not seem to be 
the case
(especially if the ``mutual friction'' in the superfluid interior
is taken into account \cite{LindM95}). 

A related question arises naturally: why are the observed masses 
of millisecond radio pulsars almost identical ? Following the standard model, 
a millisecond radio pulsar is a recycled neutron star, spun up by 
the accretion of mass and angular momentum from a companion. The observed
mass and angular velocity are those of the end of the accretion process.
Consequently the  accreted mass depends on the history of the system and 
on the nature
of the companion. By supposing ``per absurdo'' that all neutron stars are born
with the same mass, it is difficult to understand why the accreted mass is 
the {\em same} for all neutron stars. 
A possible answer is that this could result from some 
observational selection effect. For example, suppose that accreted
matter quenches the magnetic field, it is then easy to imagine that the final 
external magnetic field depends on the mass of the accreted plasma.
If the accreted mass is large enough, the magnetic field can be lower 
than the critical value for which the pulsar mechanism works.  
On the contrary, if the accreted mass is quite small,  
the magnetic field is large  and the life time of the 
radio pulsar phase is shorter and consequently more difficult to observe.

Let us conclude by saying that a lot of questions are still open about these 
systems. If we knew everything about neutron stars, such an
observation would be a waste of time. 
The only thing that we can recommend is to stay open minded. 

\subsection{CW emission from pulsars}

In this lecture, we have also investigated the CW emission resulting from the
magnetic field induced distortion of neutron stars.
The computations presented in \S~\ref{s:mag,num,res}
show that the distortion at fixed
magnetic dipole moment depends very sensitively on the magnetic configuration.  
The case of a perfect conductor interior with toroidal electric currents is the
less favorable one, even if the currents are concentrated in the crust. 
Stochastic magnetic fields (that we modeled by considering 
counter-rotating currents) enhance the deformation by 
several orders of magnitude and may lead to a detectable amplitude for 
a pulsar like the Crab. As concerns superconducting interiors --- the most
realistic configuration for neutron stars ---  we have studied  
type I superconductors numerically, with a simple magnetic structure outside 
the superconducting region. The distortion factor is then $\sim 10^2$ to $10^3$
higher than in the normal (perfect conductor) case, but still insufficient to 
lead to a positive detection by the first generation of kilometric 
interferometric detectors. We have not studied in detail the type II 
superconductor but have put forward some argument which makes it a promising
candidate for gravitational wave detection. 

\ack{We warmly thank Joachim Frieben and Pawel Haensel for their 
careful reading of the preliminary version of these notes.} 

\bigskip\goodbreak
\noindent{\bf Problem}:

Show that, at fixed angular momentum, the kinetic energy of an 
incompressible fluid in a cylinder is minimal for rigid rotation.

\bigskip
\noindent{\em Solution}:

Let $\Omega(\rho,z)$ be the angular velocity ($\rho=\sqrt{x^2+y^2})$. 
The kinetic energy of the fluid is given by 
$T=\int_V 1/2\ n\Omega^2(\rho,z)\rho^2 \, dV $ where $n$ is the 
density of the fluid (assumed to be constant). 
We have to find an extremum of this quantity under the
constraint that the angular momentum 
$L=\int_V n\, \Omega(\rho,z)\rho^2 \, dV $
has a fixed value. By means of the Lagrangian multiplier technique, this 
amounts to find an extremum of 
\be
\int_{V} {1\ov 2}\,  n\, \Omega^2(\rho,z)\rho^2 \, dV + 
\lambda\int_{V}n\, \Omega(\rho,z)\rho^2 \, dV \ , 
\ee
where $\lambda$ is the Lagrangian multiplier.
By performing the variation with respect to $\Omega$ we obtain
\be
\int_{V}(\Omega(\rho,z)\rho^2
 +\lambda \rho^2)\, \delta \Omega \,  dV =0 \ , 
\ee
from which $\Omega={\rm const}$.

\appendix
\label{a:hydro}
{\bf RELATIVISTIC HYDRODYNAMICS IN AN ACCELERATED \\ 
FRAME}

\bigskip

In this appendix, we examine how to re-write 
the equation of momentum-energy conservation
\be \label{e:div(T)=0}
    \nabla_\mu T^{\mu\alpha} = 0 
\ee
as a system of evolution equations with respect to a given observer $\cal O$,
the evolved variables being the energy density and the fluid velocity, 
both measured by $\cal O$. We consider perfect fluids only:
\be \label{e:Tab,fluide}
    T^{\alpha\beta} = (e+p) \, u^\alpha u^\beta + p\, g^{\alpha\beta} \ ,
\ee
so that the equation for the fluid velocity will constitute a relativistic
generalization of the {\em Euler equation} of classical hydrodynamics. 

The observer $\cal O$ is completely arbitrary;
he is simply described by its 4-velocity $v^\alpha$. Strictly speaking we
consider a {\em family} of observers $\cal O$ (a {\em congruence} of
worldlines), so that $v^\alpha$ constitutes a smooth vector field on 
spacetime. A fundamental tensor field related to $\cal O$
is the projection operator onto the 3-space $P$ orthogonal to $v^\alpha$:
\be
    q_{\alpha\beta} := g_{\alpha\beta} + v_\alpha\, v_\beta \ . 
\ee
The 3-space $P$ is made of spacelike vectors and can be thought as the 
``physical'' three-dimensional space ``felt'' by the observer $\cal O$. Note
that if $\cal O$ is rotating ($\omega_{\alpha\beta}\not = 0$, see below),
the vector space $P$ is not integrable in global 3-surfaces. 

The motion of the observer $\cal O$ through spacetime is characterized by
the Ehlers decomposition of $\nabla_\beta v_\alpha$ (see, e.g.,
Sect.~4.1 of ref.~\cite{HawkE73})
\be
   \nabla_\beta v_\alpha = \omega_{\alpha\beta} + \theta_{\alpha\beta}
	- a_\alpha v_\beta \ ,
\ee
where
\be
   \omega_{\alpha\beta} := q_\alpha^{\ \, \mu} q_\beta^{\ \, \nu}
	\nabla_{[\nu} \, v_{\mu]}
\ee
is the {\em rotation 2-form} of $\cal O$,
\be \label{e:expansion,tens}
  \theta_{\alpha\beta} := q_\alpha^{\ \, \mu} q_\beta^{\ \, \nu}
	\nabla_{(\nu} \, v_{\mu)} 
\ee
is the {\em expansion tensor} of $\cal O$ and
\be
  a_\alpha := v^\mu \nabla_\mu v_\alpha 
\ee
is the {\em 4-acceleration} of $\cal O$. 

For a 4-vector $W^\alpha$ lying in the ``physical space'' $P$ of $\cal O$
($v_\mu W^\mu = 0$), 
we introduce the {\em 3-covariant derivative} with respect to $\cal O$ as
\be
    \overline\nabla_\alpha W_\beta := q_\alpha^{\ \, \mu} q_\beta^{\ \, \nu}
	\nabla_\mu W_\nu \ .
\ee
This definition results in the following relation between the 4-covariant
and the 3-covariant derivatives:
\be
   \nabla_\alpha W_\beta = \overline\nabla_\alpha W_\beta - v_\alpha\, 
	v^\mu \nabla_\mu W_\beta + ( \theta_{\alpha\mu} W^\mu - 
	\omega_{\alpha\mu} W^\mu ) v_\beta \ ,
\ee
from which the following relation between the 4-divergence and the
3-divergence is immediately derived:
\be
	\nabla_\mu W^\mu = \overline\nabla_\mu W^\mu + a_\mu W^\mu \ .
\ee

The fluid motion as seen by the observer $\cal O$ is specified by the
{\em Lorentz factor}
\be
    \Gamma := - v_\mu u^\mu \ ,
\ee
and the {\em 3-velocity}
\be
    V^\alpha := {1\ov \Gamma} q^\alpha_{\ \, \mu} u^\mu \ .
\ee
$V^\alpha$ belongs to $P$ and 
is the fluid velocity as measured by the observer $\cal O$ with 
his clock and his ruler. 
The following relations are immediate consequences of the above definitions:
\begin{eqnarray}
   u^\alpha & = & \Gamma (V^\alpha + v^\alpha) \label{e:u(V,v)} \\
   \Gamma & = & (1-V_\mu V^\mu)^{-1/2} \ .
\end{eqnarray}
The fluid energy density as measured by $\cal O$ is given by the formula
\be
    E = T_{\mu\nu} \, v^\mu v^\nu \ ,
\ee
or, according to the form (\ref{e:Tab,fluide}) of $T_{\mu\nu}$, 
\be \label{e:E(e,p)}
     E = \Gamma^2 (e+p) - p \ .
\ee
From this expression, it is clear that the kinetic energy of the fluid with 
respect to $\cal O$ is included in the energy density $E$ via the Lorentz
factor $\Gamma$. 

Having set these definitions, let us now examine the equations of motion 
deduced from the momentum-energy conservation, Eq.~(\ref{e:div(T)=0}), 
which, using the perfect fluid form (\ref{e:Tab,fluide}) of 
$T_{\alpha\beta}$, can be written as 
\be \label{e:div(T),u}
  (e+p) u^\mu \nabla_\mu u^\alpha + 
  \nabla_\mu \l[ (e+p) u^\mu \r] \, u^\alpha + \nabla^\alpha p = 0
\ee 

The evolution equation for the fluid energy density $E$ as measured by
$\cal O$ is obtained by projecting Eq.~(\ref{e:div(T),u}) along $v^\alpha$. 
Invoking Eqs.~(\ref{e:u(V,v)}) and (\ref{e:E(e,p)}), one obtains
after straightforward calculations
\be \label{e:evol,E}
   v^\mu \nabla_\mu E + \overline\nabla_\mu \l[ (E+p) V^\mu \r] 
	+ (E+p) ( 2 a_\mu V^\mu + \theta_\mu^{\ \, \mu} 
	+ \theta_{\mu\nu} V^\mu V^\nu ) = 0 \ .
\ee

The relativistic generalization of the Euler equation is obtained by 
projecting Eq.~(\ref{e:div(T),u}) onto $P$, by means of $q_{\alpha\beta}$. 
After straightforward calculations (at a certain stage, use must be made of
Eq.~(\ref{e:evol,E})) one obtains
\begin{eqnarray}
  & & v^\mu \nabla_\mu V^\alpha - a_\mu V^\mu v^\alpha + 
  V^\mu \overline\nabla_\mu V^\alpha + (\omega^\alpha_{\ \, \mu} + 
	\theta^\alpha_{\ \, \mu} ) V^\mu 
  - (a_\mu V^\mu + \theta_{\mu\nu} V^\mu V^\nu) V^\alpha \nonumber \\
 & & \qquad = - {1\ov E+p} \l( \overline\nabla^\alpha p + 
	V^\alpha v^\mu \nabla_\mu p \r) - a^\alpha  \label{e:Euler,vdV} \ . 
\end{eqnarray}
Note that the first two terms on the left-hand side,
$v^\mu \nabla_\mu V^\alpha  - a_\mu V^\mu v^\alpha$, constitute the 
{\em Fermi-Walker derivative} \cite{HawkE73} of $V^\alpha$ with respect to 
$v^\alpha$.  
The Fermi-Walker derivative measures the rate of change {\em within} $P$ 
of $V^\alpha$ with respect to the proper time of $\cal O$.  
If the observer $\cal O$ has set up a local coordinate system, with respect
to which the length of the vectors in $P$ are evaluated, a derivative
operator more convenient than $v^\mu \nabla_\mu$ is the {\em Lie derivative}
along $v^\alpha$, $\pounds_v$. The term which naturally appears on the
left-hand side of Eq.~(\ref{e:Euler,vdV}) is then the 
{\em convected derivative} of $V^\alpha$ \cite{CartQ72}, \cite{Carte80} :
\be
    D_v V^\alpha := \pounds_v V^\alpha - a_\mu V^\mu v^\alpha = 
	v^\mu \nabla_\mu V^\alpha - (\omega^\alpha_{\ \, \mu} +
	\theta^\alpha_{\ \, \mu}) V^\mu - a_\mu V^\mu v^\alpha \ .
\ee
The Euler equation (\ref{e:Euler,vdV}) then becomes 
\begin{eqnarray}
  & & D_v V^\alpha + 
  V^\mu \overline\nabla_\mu V^\alpha + 2(\omega^\alpha_{\ \, \mu} + 
	\theta^\alpha_{\ \, \mu} ) V^\mu 
  - (a_\mu V^\mu + \theta_{\mu\nu} V^\mu V^\nu) V^\alpha \nonumber \\
 & & \qquad = - {1\ov E+p} \l( \overline\nabla^\alpha p + 
	D_v p \, V^\alpha \r) - a^\alpha  \label{e:Euler,DV} \ . 
\end{eqnarray}
In the Newtonian limit, $D_v$ reduces simply to $\partial/\partial t$. 
Moreover, the $V^\mu \overline\nabla_\mu V^\alpha$ gives 
the classical term $(\vec V \cdot \vec\nabla)\vec V$ and
$2\, \omega^\alpha_{\ \, \mu} V^\mu$ gives the Coriolis term
$2\, \vec\omega\times\vec V$, induced by the
rotation of the observer $\cal O$ with respect to some inertial frame.   
The terms involving $\theta_{\alpha\beta}$ are due to the non-rigidity
of the frame set up by the observer $\cal O$. On the right-hand side,
the classical $(1/\rho) \vec\nabla p$ term is recognized, 
supplemented by the special relativistic term 
$(\partial p/\partial t) \, \vec V$. The last term, 
the acceleration $-a^\alpha$,
contains gravitational as well as centrifugal forces.


\begin{thebibliography}{100}

\bibitem{Blanc96} Blanchet L., this volume.

\bibitem{Haens96} Haensel P., this volume.

\bibitem{Chand70} Chandrasekhar S.,
\review Phys. Rev. Lett., 24, 1970, 611.

\bibitem{FrieS78} Friedman J.L., Schutz B.F.,
\review Astrophys. J., 222, 1978, 281.

\bibitem{Fried78} Friedman J.L.,
\review Commun. Math. Phys., 62, 1978, 247.

\bibitem{RobeS63} Roberts P.H., Stewartson K.,
\review Astrophys. J., 137, 1963, 777.

\bibitem{Chand69} Chandrasekhar S.,
\book Ellipsoidal figures of equilibrium, Yale University Press, New Haven,
1969, 1.

\bibitem{Tasso78} Tassoul J.-L.,
\book Theory of rotating stars, Princeton University Press, Princeton,
1978, 1.

\bibitem{ChrCK96} Christodoulou D.M., Contopoulos J., Kazanas D.,
Interchange method
in incompressible magnetized Couette flow: structural and magnetorotational
instabilities, preprint.

\bibitem{LaiRS93} Lai D., Rasio F.A., Shapiro S.L.,
\review Astrophys. J. Suppl., 88, 1993, 205.

\bibitem{PresT73} Press W.H., Teukolsky S.A.,
\review Astrophys. J., 181, 1973, 513.

\bibitem{Mille74} Miller B.D.,
\review Astrophys. J., 187, 1974, 609.

\bibitem{LandL76} Landau L.D., E. Lifchitz,
\book Physique statistique, Mir, Moscow, 1984, 488.

\bibitem{BertR76} Bertin G., Radicati L.A.,
\review Astrophys. J., 206, 1976, 815. 

\bibitem{ChKST95a} Christodoulou D.M., Kazanas D., Shlosman I.,
Tohline J.E., \review Astrophys. J., 446, 1995, 472.

\bibitem{ChKST95b} Christodoulou D.M., Kazanas D., Shlosman I.,
Tohline J.E., \review Astrophys. J., 446, 1995, 485.

\bibitem{ChKST95c} Christodoulou D.M., Kazanas D., Shlosman I.,
Tohline J.E., \review Astrophys. J., 446, 1995, 500.

\bibitem{ChKST95d} Christodoulou D.M., Kazanas D., Shlosman I.,
Tohline J.E., \review Astrophys. J., 446, 1995, 510.

\bibitem{Jeans19} Jeans J.H.,
\book Problems of cosmogony and stellar dynamics,
Cambridge University Press, Cambridge, 1919, 1.

\bibitem{Jeans28} Jeans J.H.,
\book Astronomy and cosmogony, Cambridge University
Press, Cambridge, 1928, 1.
\ Reprinted in 1961 (Dover, New York).

\bibitem{James64} James R.A.,
\review Astrophys. J., 140, 1964, 552.

\bibitem{IpseM85} Ipser J.R., Managan R.A.,
\review Astrophys. J., 292, 1985, 517.

\bibitem{Manag85} Managan R.A.,
\review Astrophys. J., 294, 1985, 463.

\bibitem{ImaFD85} Imamura J.N., Friedman J.L., Durisen R.H.,
\review Astrophys. J., 294, 1985, 474.

\bibitem{Lindb95} Lindblom L.,
\review Astrophys. J., 438, 1995, 265.

\bibitem{YoshE95} Yoshida S., Eriguchi Y.,
\review Astrophys. J., 438, 1995, 830.

\bibitem{FrieI92} Friedman J.L, Ipser J.R.,
\review Phil. Trans. R. Lond. A, 340, 1992, 391.
\ Reprinted in: Classical general relativity, ed. S.~Chandrasekhar
(Oxford University Press, Oxford, 1993).

\bibitem{LindM95} Lindblom L., Mendell G.,
\review Astrophys. J., 444, 1995, 804.

\bibitem{IpseM84} Ipser J.R., Managan R.A.,
\review Astrophys. J., 282, 1984, 287.

\bibitem{Wagon84} Wagoner R.W.,
\review Astrophys. J., 278, 1984, 345.

\bibitem{LaiSh95} Lai D., Shapiro S.L.,
\review Astrophys. J., 442, 1995, 259.

\bibitem{LindD77} Lindblom L., Detweiler S.L.,
\review Astrophys. J., 211, 1977, 565.

\bibitem{BonFG96} Bonazzola S., Frieben J., Gourgoulhon E.,
\review Astrophys. J., 460, 1996, 379. 

\bibitem{Carte79} Carter B.,
in Active Galactic Nuclei, eds. C. Hazard \& S. Mitton
(Cambridge University Press, Cambridge, 1979).

\bibitem{BoGSM93} Bonazzola S., Gourgoulhon E., Salgado M., Marck J.-A.,
\review Astron. Astrophys., 278, 1993, 421.

\bibitem{Lichn67} Lichnerowicz A.,
\book Relativistic hydrodynamics and magnetohydrodynamics, 
W.A. Benjamin Inc., New York, 1967, 1.

\bibitem{Carte73} Carter B.,  1973,
in Black Holes --- Les Houches 1972, eds.
C. DeWitt \& B.S. DeWitt (Gordon \& Breach, New York, 1973).

\bibitem{AbraY96} Abrahams A.M., York J.W., this volume. 

\bibitem{Barde70} Bardeen J.M.,
\review Astrophys. J., 162, 1970, 71.

\bibitem{Geroc71} Geroch R.,
\review J. Math. Phys., 12, 1971, 918.

\bibitem{Gourg96} Bonazzola S., Gourgoulhon E., Marck J.-A., this volume.

\bibitem{SaBGH94} Salgado M., Bonazzola S., Gourgoulhon E., Haensel P.,
\review Astron. Astrophys., 291, 1994, 155.

\bibitem{IpseM81} Ipser J.R., Managan R.A.,
\review Astrophys. J., 250, 1981, 362.

\bibitem{HachE82} Hachisu I., Eriguchi Y.,
\review Prog. Theor. Phys., 68, 1982, 206.

\bibitem{Pandh71} Pandharipande V.R.,
\review  Nucl. Phys. A, 174, 1971, 641. 

\bibitem{BethJ74} Bethe H.A., Johnson M.B.,
\review Nucl. Phys. A, 230, 1974, 1.

\bibitem{FrieP81} Friedman B., Pandharipande V.R.,
\review  Nucl. Phys. A, 361, 1981, 502. 

\bibitem{HaeKP81} Haensel P., Kutschera M., Pr\'oszy\'nski M.,
\review Astron. Astrophys., 102, 1981, 299.

\bibitem{Diaz85} Diaz Alonso J.,
\review Phys. Rev. D, 31, 1985, 1315.

\bibitem{Glend85} Glendenning N.K., 
\review Astrophys. J., 293, 1985, 470.

\bibitem{WirFF88} Wiringa R.B., Fiks V., Fabrocini A.,
\review Phys. Rev. C, 38, 1988, 1010.

\bibitem{WebGW91} Weber F., Glendenning N.K., Weigel M.K.,
\review Astrophys. J., 373, 1991, 579.

\bibitem{Thorn80} Thorne K.S.,
\review Rev. Mod. Phys., 52, 1980, 299.

\bibitem{Ipser71} Ipser J.R.,
\review Astrophys. J., 166, 1971, 175.

\bibitem{BonaG96} Bonazzola S., Gourgoulhon E.,
{\sl Astron. Astrophys.}, in press (preprint: {\sl astro-ph/9602107}). 

\bibitem{ArneB77} Arnett W.D., Bowers R.L.,
\review Astrophys. J. Suppl., 33, 1977, 415.

\bibitem{NeCJT95} New K.C.B., Chanmugam G., Johnson W.W., Tohline J.E.,
\review Astrophys. J., 450, 1995, 757. 

\bibitem{Strau84} Straumann N., 
\book General relativity and relativistic astrophysics, Springer
Verlag, Berlin, 1984, 353. 

\bibitem{BoBGN95} Bocquet M., Bonazzola S., Gourgoulhon E., Novak J., 
\review Astron. Astrophys., 301, 1995, 757.

\bibitem{TayML93} Taylor J.H., Manchester R.N., Lyne A.G.,
\review Astrophys. J. Suppl., 88, 1993, 529.

\bibitem{TaMLC95} Taylor J.H., Manchester R.N., Lyne A.G., Camilo F., 
unpublished work (1995). 

\bibitem{BrilG96} Brillet A., Giazotto A., this volume. 

\bibitem{Bondu96} Bondu F., PhD Thesis (Universit\'e Paris XI, 1996).  

\bibitem{ThomD93} Thompson C., Duncan R.C., 
\review Astrophys. J., 408, 1993, 194. 

\bibitem{Ruder91} Ruderman M., 
\review Astrophys. J., 382, 1991, 576. 

\bibitem{Mulle96} M\"uller E., this volume. 

\bibitem{ThAKT93} Thorsett S.E., Arzoumanian Z., McKinnon M.M., Taylor J.H.,
\review Astrophys. J., 405, 1993, L29. 
  
\bibitem{Wolsz96} Wolszczan A., this volume. 

\bibitem{Lindb96} Lindblom L., private communication. 

\bibitem{HawkE73} Hawking S.W., Ellis G.F.R.,
\book The large scale structure of space-time, 
Cambridge University Press, Cambridge, 1973, 1.

\bibitem{CartQ72} Carter B., Quintana H.,
\review Proc. R. Soc. Lond. A, 331, 1972, 57.

\bibitem{Carte80} Carter B.,
\review Proc. R. Soc. Lond. A, 372, 1980, 169.

\end{thebibliography}
\end{document}